\documentclass{emulateapj}
\usepackage{savesym}
\savesymbol{tablenum}
\usepackage{siunitx}
\restoresymbol{SIX}{tablenum}
\let\oldendlongtable\endlongtable
\let\oldlongtable\longtable
\makeatletter
\renewenvironment{longtable}[1]{
\oldlongtable{[#1]}
\def\pt@format{#1}
}{\oldendlongtable}
\makeatother

\usepackage{color}
\definecolor{white}{rgb}{1,1,1}

\usepackage{colortbl}
\definecolor{lmugreen}{rgb}{0.01,0.58,0.25}

\usepackage{amsmath}
\usepackage{natbib}
\usepackage{float}
\usepackage{dcolumn}
\usepackage{booktabs}
\usepackage[
	colorlinks=true,
	urlcolor=lmugreen,
	linkcolor=lmugreen,
	citecolor=lmugreen
	]{hyperref}
\usepackage{rotating}
\usepackage{graphicx}
\usepackage{tablefootnote}

\usepackage{longtable}
\usepackage[nolist]{acronym}

\usepackage{listings}
\newcommand{\citeorder}[1]{}

\defcitealias{KDR1:inprep}{Paper I}
\defcitealias{KDR2a:inprep}{Paper II}
\defcitealias{KDR2b:inprep}{Paper III}
\defcitealias{KDR3:inprep}{Paper IV}
\defcitealias{KoepferlRobitaille:inprep}{Paper FluxCompensator}

\DeclareSIUnit\yr{yr}
\DeclareSIUnit\ergs{ergs}
\DeclareSIUnit\Jy{Jy}
\DeclareSIUnit\Hz{Hz}
\DeclareSIUnit\sr{sr}
\DeclareSIUnit\pc{pc}
\DeclareSIUnit\ppm{ppm}
\DeclareSIUnit\kpc{\kilo\pc}
\DeclareSIUnit\AU{AU}
\DeclareSIUnit\eqsign{\text{\ensuremath{=}}}
\DeclareSIUnit\simeqsign{\text{\ensuremath{\simeq}}}
\DeclareSIUnit\mag{mag}
\DeclareSIUnit\gramms{g}
\DeclareSIUnit\gccm{\gramms\per\cubic\centi\metre}
\DeclareSIUnit\gcm{\gramms\per\square\centi\metre}
\DeclareSIUnit\microns{\micro\metre}
\DeclareSIUnit\Myr{\mega\yr}
\DeclareSIUnit\Msun{\text{\ensuremath{M_\odot}}}
\DeclareSIUnit\Lsun{\text{\ensuremath{L_\odot}}}
\DeclareSIUnit\Rsun{\text{\ensuremath{T_\odot}}}
\DeclareSIUnit\Minfall{\Msun\per\yr}

\slugcomment{The Astrophysical Journal}

\shorttitle{Synthetic Star-forming Regions - II. Verifying Dust \& Gas Measurements with Modified Blackbody Fitting}
\shortauthors{Koepferl, Robitaille, Dale}

\begin{document}

\title{Insights from Synthetic Star-forming Regions: \\II. Verifying Dust Surface Density, Dust Temperature \& Gas Mass Measurements \\with Modified Blackbody Fitting}
\author{Christine M. Koepferl$^{1,2}$, Thomas P. Robitaille$^{1,3}$, and James E. Dale$^4$}
\affil{$^1$ Max Planck Institute for Astronomy, K\"onigstuhl 17, D-69117 Heidelberg, Germany\\
$^2$ Scottish Universities Physics Alliance (SUPA), School of Physics and Astronomy, University of St Andrews\\North Haugh, St Andrews, KY16 9SS, UK\\
$^3$ Freelance Consultant, Headingley Enterprise and Arts Centre, Bennett Road Headingley, Leeds LS6 3HN\\
$^4$ University Observatory Munich, Scheinerstr. 1, D-81679 Munich, Germany}
\email{cmk8@st-andrews.ac.uk}
\received{29 January 2016}
\accepted{23 June 2016}

\begin{abstract}
We use a large data-set of realistic synthetic observations (produced in \citetalias{KDR1:inprep}) to assess how observational techniques affect the measurement physical properties of star-forming regions. In this part of the paper series \citepalias{KDR2a:inprep}, we explore the reliability of the measured total gas mass, dust surface density and dust temperature maps derived from modified blackbody fitting of synthetic \emph{Herschel} observations. We found from our pixel-by-pixel analysis of the measured dust surface density and dust temperature a worrisome error spread especially close to star-formation sites and low-density regions, where for those "contaminated" pixels the surface densities can be under/over-estimated by up to three orders of magnitude. In light of this, we recommend to treat the pixel-based results from this technique with caution in regions with active star formation. In regions of high background typical in the inner Galactic plane, we are not able to recover reliable surface density maps of individual synthetic regions, since low-mass regions are lost in the \acs{FIR} background.
When measuring the total gas mass of regions in moderate background, we find that modified blackbody fitting works well (absolute error: +\SI{9}{\percent}; \SI{-13}{\percent}) up to \SI{10}{\kpc} distance (errors increase with distance). Commonly, the initial images are convolved to the largest common beam-size, which smears contaminated pixels over large areas. The resulting information loss makes this commonly-used technique less verifiable as now $\chi^2$-values cannot be used as a quality indicator of a fitted pixel. Our control measurements of the total gas mass (without the step of convolution to the largest common beam size) produce similar results (absolute error: +\SI{20}{\percent}; \SI{-7}{\percent}) while having much lower median errors especially for the high-mass stellar feedback phase. In upcoming papers \citepalias{KDR2b:inprep,KDR3:inprep} of this series we test the reliability of measured \acl{SFR} with direct and indirect techniques.
\end{abstract}

\section{Introduction}
\label{C5:Sec:Introduction}
The gas in molecular clouds sets the scene for the star formation process and is the initial ingredient of stars. The fundamental properties of the gas and dust are the total mass $M_{\textup{gas}}$, the dust surface density $\Sigma_{\rho}^{\textup{dust}}$, the dust temperature $T_{\textup{dust}}$ and the dust-to-gas ratio. All these parameters are in some sense essential to recover information about the star-formation process. For instance, the different morphological features of the gas and dust in star-forming regions can be studied from the 2-d projection maps (e.\,g.~$\Sigma_{\rho}^{\textup{dust}}$) and also through log-scaled column density histograms, referred to as column
density probability distribution functions (\acs{N-PDF}). The \acsp{N-PDF} are constructed from dust surface density maps $\Sigma_{\rho}^{\textup{dust}}$. For more details, see the reviews of \citealt{Andre:PPVI:2014} and \citealt{Padoan:2014} and also examples for observational application by \cite{Kainulainen:2013} or \cite{Lombardi:2015}. 

Recently, column density maps have also been compared to hydrodynamical simulations (e.\,g.~\citealt{Roccatagliata:2014}, \citealt{SmithRowan:fila:2014}, \citealt{SmithRowan:fila:2013}), where also substructures such as filaments and cores can be studied.

\subsection{Dust \& Gas Property Measurement Techniques}
The fundamental properties of the gas and dust can be calculated in several different ways:

\begin{itemize}
\item \textbf{Line Gas Tracers}\\
    The gas properties in a region can be calculated directly through techniques that make use of gas tracers. For instance, the total gas mass $M_{\textup{gas}}$ can be measured dynamically by relating observed spectral line widths to virial masses. Volume densities can be estimated when observing spectral line transitions that are primarily stimulated in dense regions above a certain density threshold. For more information about techniques which measure the gas properties directly, see the reviews of \cite{Bolatto:2013} and \cite{Bergin:2007} as well as the work by \cite{Shirley:2015} and \cite{BertoldiMcKee:1992} (Appendix).
\item \textbf{Optical Depth}\\
            The total gas mass $M_{\textup{gas}}$ of close-by regions can be calculated from the measured optical depth $\tau_\nu$ of a star-forming region. With the optical depth $\tau_\nu$, the dust column density $\Sigma_{\rho}^{\textup{dust}}$ can be extracted and converted to a gas column density $\Sigma_{\rho}^{\textup{gas}}$ using an assumed dust-to-gas ratio. When the distance $D$ is known, the total gas mass $M_{\textup{gas}}$ can also be recovered. 
            Optical depth $\tau_\nu$ of a region can be measured in several ways. One option is to recover the extinction level from the surface density of stars \citep[for a description see][]{Cambresy:2002}; as fewer stars can be seen in the stellar field when inter-stellar material blocks (extincts) their light towards the observer. In the extincted region one sees less stars in the stellar background. With the recovered extinction maps $A_\nu$ (e.\,g.~from the $K$ band) the optical depth $\tau_\nu$ can be recovered ($\tau_\nu=A_\nu/1.086$, see \citealt{Carroll:2014,Carroll:1996}). Alternatively, the optical depth $\tau_\nu$ can be estimated from the color excess of the background stars \citep[e.\,g.][]{KainulainenTan:2013,Lombardi:2009,LombardiAlves:2001,Lada:1994}. 

\item \textbf{Continuum Dust Tracers}\\
    The gas properties of a region can also be estimated indirectly through the thermal emission of the dust species. The dust surface density $\Sigma_{\rho}^{\textup{dust}}$ can be inferred from a single continuum wavelength under the assumption of a dust temperature $T_{\textup{dust}}$ and assuming the dust is optically thin. This technique, described by \cite{Hildebrand:1983}, also allows one to infer the total gas mass $M_{\textup{gas}}$ when assuming a dust-to-gas ratio. For observations of the dust emission in several different bands, the gas mass $M_{\textup{gas}}$, as well as the dust temperature $T_{\textup{dust}}$, can be recovered using modified blackbody fitting as long as the following \textit{assumptions} hold:
    \begin{itemize}
    \item constant temperature along the line of sight\\[-0.5cm]
    \item absence of scattering\\[-0.5cm]
    \item constant dust-to-gas ratio\\[-0.5cm]
    \item constant dust properties such as the emissivity
    \end{itemize}
    This technique which makes use of several dust continuum observations at different wavelengths is known as greybody fitting or modified blackbody fitting. A detailed study of how a non-constant line-of-sight temperature can affect the properties extracted from the modified blackbody fitting technique was presented by \cite{Shetty:2009} especially when the observed object is above their estimated temperature threshold.
\end{itemize}

Since \emph{Herschel} observations have become available, the physical properties of the gas/dust have been measured with the \textit{modified blackbody fitting technique} by the Galactic as well as the extra-galactic communities \citep[e.\,g.][]{Stutz:2015,Battersby:2014,Roccatagliata:2014,Sadavoy:2014,Launhardt:2013,Galametz:2012,Groves:2012,SmithM:2012,Andre:2010,Elbaz:2010,Konyves:2010,Stutz:2010}. 

Recently, several variations of modified blackbody fitting have been published, for instance by \cite{Kelly:2012} where a Bayesian framework is used to infer the surface density and temperature. Additionally, \cite{Marsh:2015} developed a technique, which is different to the commonly used approach of modified blackbody fitting where all bands are used simultaneously to reach a result. The \cite{Marsh:2015} technique makes use of every band individually to produce channel maps of surface densities for different temperatures while adopting a Bayesian approach. Both studies produced results in agreement with the results from the commonly used standard modified blackbody fitting technique.

\subsection{Motivation}
The measured observable properties of the gas and dust (total gas mass $M_{\textup{gas}}$, dust surface density $\Sigma_{\rho}^{\textup{dust}}$, dust temperature $T_{\textup{dust}}$) of star-forming regions are essential for the research field of star formation. For instance, they are used to test theoretical simulations, which might then help us understand the main driving mechanisms of star formation. For this reasons, it is beneficial that the measurements of these properties are as accurate as possible. Therefore, we will test the accuracy of the standard \textit{modified blackbody fitting technique} which is commonly used to infer these physical gas and dust properties of star-forming regions. We especially focus on the reliability of the technique on global scales for distances between \SI{3}{\kpc} and \SI{10}{\kpc}. We achieve this with the large set of realistic synthetic observations developed in \cite{KDR1:inprep} referred to as \citetalias{KDR1:inprep}. 

\subsection{Outline}
With these realistic synthetic observations, which are directly comparable to real observations (recapitulation in Section~\ref{C5:Sec:recap}), we can start our analysis of star-formation properties measured with dust tracers. In this paper, hereafter referred to as \citetalias{KDR2a:inprep}, we will use modified blackbody fitting to estimate the dust surface density, the dust temperature and the total gas mass in the synthetic cloud for different time-steps, orientations and distances and will discuss the accuracy of the technique. We will describe the formalism of the tested technique in Section~\ref{C5:Sec:Mgas_MBBF}, the assumptions in Section~\ref{C5:Sec:Mgas_method} and the results in Section~\ref{C5:Sec:Mgas_Results}. In Section~\ref{C5:Sec:Discussion}, we discuss the biases of the technique before we summarize our findings in Section~\ref{C5:Sec:Summary}. In follow-up papers \cite{KDR2b:inprep} and \citeauthor[(in prep.)]{KDR3:inprep}, referred to as \citetalias{KDR2b:inprep} and \citetalias{KDR3:inprep}, we will explore the reliability of different diffuse star-formation dust tracers and direct counting techniques to calculate the \ac{SFR}.

\section{Recapitulation}
\label{C5:Sec:recap}
To summarize \citetalias{KDR1:inprep}, we used \ac{SPH} simulations of a synthetic star-forming region at several time-steps to produce synthetic observations. We chose the \ac{SPH} simulations of \cite{DaleI:2011} and \cite{DaleIoni:2012,DaleIoni:2013,DaleWind:2013,DaleBoth:2014}, referred to as \acs{D14}, because they include high-mass stellar feedback, such as stellar ionization and winds, and cover relatively large scales (\textit{run I}: \SI{e4}{\Msun} in mass, about \SI{30}{\pc} in diameter). For more details about the \ac{SPH} simulations used, see \citetalias{KDR1:inprep} or \acs{D14}. The evolution of the properties over time is displayed in Table~1 of \citetalias{KDR1:inprep}.

We extend (post-process) the \ac{SPH} simulations with radiative transfer calculations, which account for the stellar heating of the dust. We used \textsc{Hyperion}, a 3-d dust continuum Monte-Carlo radiative transfer code. For more details about \textsc{Hyperion}, see \cite{Robitaille:2011} and radiative transfer in general, see \cite{Steinacker:2013}.

In \citetalias{KDR1:inprep}, we developed algorithms to map the particle-based \ac{SPH} simulations onto a Voronoi mesh for the radiative transfer, while conserving the mass. We decided to refine the density structure near the accreting protostars beyond the resolution limit of the \ac{SPH} simulation in order to recover the \ac{MIR} flux better. For this inward extrapolation, we explored both a rotationally flattened \citep{Ulrich:1976} envelope profile and a power-law envelope description. For computational efficiency, we decided to use precomputed analytical models of \acp{YSO} in the radiative transfer calculation on the smallest scales ($<\SI{500}{\AU}$). The stellar objects are set up with fitted properties from pre-main-sequence tracks \citep{Siess:2000, Bernasconi:1996} and photosphere models \citep{Castelli:2004, Brott:2005}. The number of stellar objects in every time-step can be followed in Table~1 of \citetalias{KDR1:inprep}.

We assumed a dust-to-gas ratio of $\frac{\textup{dust}}{\textup{gas}}=\num{0.01}$ \citep{DraineBook} and used \cite{Draine:2007} dust grain properties for the radiative transfer calculations. 
Furthermore, we coupled the radiative transfer dust temperature with an ambient background temperature $T_{\textup{iso}}=\SI{18}{\kelvin}$, typical for relatively empty patches of the Galactic plane (see also Section~\ref{C5:Sec:Mgas_Results}). For more information about the radiative transfer set-up of synthetic star-forming regions, see \citetalias{KDR1:inprep}.

We produced realistic synthetic observations for all the radiative transfer images using the \textsc{FluxCompensator} (\citetalias{KoepferlRobitaille:inprep}: \citeauthor[][in prep.]{KoepferlRobitaille:inprep}). This package accounts for the extinction, the transmission curves of the telescope and detector, the pixel-size, the \ac{PSF} convolution and produces realistic images. Further, a realistic background can be generated by combining the realistic synthetic observations with a relatively empty patch of the Galactic plane. We produced each image twice: without and with combined background (see also Section~\ref{C5:Sec:Mgas_Results}). In total, we produced about 5800 realistic synthetic observations of the following configurations with the \textsc{FluxCompensator}, which are provided in the online material of \citetalias{KDR1:inprep}:
\begin{itemize}
\item {\bf 23 Time-steps}\\
We follow 23 equally spaced time-steps over \SI{3.3}{\Myr} from the formation of the first star until roughly when the first supernova would goes off. The step width is $\Delta t=\SI{149000}{\yr}$. Ionization and winds of high-mass stars are switched on once three high-mass stellar particles above \SI{20}{\Msun} have formed. Here this happens approximately \SI{1.7}{\Myr} after the first star has formed.
\vspace*{-0.15cm}
\item {\bf 3 Circumstellar Set-ups}\\
We use three different circumstellar set-ups: two envelope refinements beyond the resolution threshold of the simulation and one without added envelopes (\acs{CM1}), as a control run to see whether refinement is necessary. The two envelope refinement set-ups we used were a rotationally flattened envelope \citep[\acs{CM2},][]{Ulrich:1976} and a power-law envelope (\acs{CM3}). In both cases, we include a circumstellar disk through a pre-computed analytical model.
\vspace*{-0.15cm}
\item {\bf 3 Orientations}\\
We produced synthetic observations for three mutually perpendicular viewing angles: \ac{O1}, \ac{O2} and \ac{O3}.
\vspace*{-0.15cm}
\item {\bf 2 Distances}\\
The synthetic observations are placed at two distances. To compare to nearby high-mass star-forming regions, such as Carina, Westerhout 4,5 and the Eagle Nebula, we used \SI{3}{\kpc} (\acs{D1}) and for star-forming regions across the Galactic plane we used \SI{10}{\kpc} (\acs{D2}). We account for interstellar extinction with $A_V=\num{10}$ and $A_V=\num{20}$ respectively, using the extinction law from \cite{Kim:1994} and the built-in function of the \textsc{FluxCompensator}.
\vspace*{-0.15cm}
\item {\bf 7 Bands}\\
Realistic synthetic observations in the following bands have been produced using the appropriate transmission curve, the appropriate \ac{PSF} and the appropriate pixel-size: \ac{IRAC} \SI{8}{\microns} (\ang{;;1.98}, \ang{;;1.2}), \ac{MIPS} \SI{24}{\microns} (\ang{;;6.0}, \ang{;;2.4}), \ac{PACS} \SI{70}{\microns} (\ang{;;4.4}, \ang{;;3.2}), \ac{PACS} \SI{160}{\microns} (\ang{;;9.9}, \ang{;;4.5}), \ac{SPIRE} \SI{250}{\microns} (\ang{;;17.6}, \ang{;;6.0}), \ac{SPIRE} \SI{350}{\microns} (\ang{;;23.9}, \ang{;;8.0}) and \ac{SPIRE} \SI{500}{\microns} (\ang{;;35.2}, \ang{;;11.5}) with (\acs{FWHM}, pixel-size) in brackets. The resulting images with band specific pixel resolution and beam size are labeled resolution version \acs{R1}.
\vspace*{-0.15cm}
\item {\bf 2 Backgrounds}\\
 We constructed every synthetic observation of the above configurations twice, once without (\acs{B1}) and once combined with a realistic background (\acs{B2}). For the remainder of this paper, we will test the technique for the two different background versions individually (see Section~\ref{C5:Sec:Mgas_Results} and Section~\ref{C5:Sec:results_B2}).
\end{itemize}

For more information about the construction of these realistic synthetic observations, see \citetalias{KDR1:inprep} and \citetalias{KoepferlRobitaille:inprep}.

\section{Formalism | Modified Blackbody Fitting}
\label{C5:Sec:Mgas_MBBF}
In what follows, we will describe the derivation of this technique starting off with the radiative transfer equations \citep[see][]{Carroll:2014,Carroll:1996}. In the absence of background radiation $I_{\nu,0}$, and when the source function $S_\nu$ is constant, the solution of the observed surface brightness $\Sigma_{B,\nu}$ of a cloud with optical depth $\tau_\nu$ simplifies to:
\begin{eqnarray}
    \label{C1:Eq:RT_Solution_S_const_no_background}
    \Sigma_{B,\nu}(\tau_\nu) &=& S_\nu\left(1- \exp(-\tau_\nu)\right).
\end{eqnarray}
Under the assumption of \ac{LTE}, the source function $S_\nu$ is equal to the Planck function $B_\nu(T_{\textup{dust}})$:
\begin{eqnarray}
    \label{C1:Eq:LTE_Mgas}
\Sigma_{B,\nu}(\tau_\nu) &=& B_\nu(T_{\textup{dust}})\left(1- \exp(-\tau_\nu)\right)
\end{eqnarray}
Eq.~\ref{C1:Eq:LTE_Mgas} can be simplified further if we assume that the region is optically thin:
\begin{eqnarray}
    \label{C1:Eq:LTE_Mgas_optThin}
\Sigma_{B,\nu}(\tau_\nu\ll1) = B_\nu(T_{\textup{dust}})\tau_\nu.
\end{eqnarray}
We can thus derive a relation between the observed brightness $\Sigma_{B,\nu}$ and the dust surface density $\Sigma_{\rho}^{\textup{dust}}$:
\begin{eqnarray}
    \label{C1:Eq:BBfitting1}
\Sigma_{B,\nu}(\tau_\nu\ll1) = B_\nu(T_{\textup{dust}}) \kappa_\nu \Sigma_\rho^{\textup{dust}} 
\end{eqnarray}
For common units, Eq.~\ref{C1:Eq:BBfitting1} takes the form:
\begin{eqnarray}
    \label{C5:Eq:BBfitting}
    \frac{\Sigma_{B,\nu}}{\si{\Jy\per\sr}}&=&  10^{23}\left(\frac{\Sigma_\rho^{\textup{dust}}}{\si{\gram\per\centi\meter\squared}}\right) \left(\frac{\kappa_\nu}{\si{\centi\meter\squared\per\gram}}\right)\times\\ &&\left(\frac{B_\nu(T_{\textup{dust}})}{\si{\ergs\per\second\per\centi\meter\squared\per\hertz\per\sr}}\right).\nonumber
\end{eqnarray}
Now, with these equations and a set of \ac{FIR} brightness observations $\Sigma_{B,\nu}$ (of equal pixel size and a certain frequency $\nu$) the dust surface density $\Sigma_\rho^{\textup{dust}}$ and the dust temperature $T_{\textup{dust}}$ can be estimated. Using the observations, we fit the function given by Eq.~\ref{C5:Eq:BBfitting} for every pixel, solving for the unknowns $T_{\textup{dust}}$ and $\Sigma_\rho^{\textup{dust}}$ and recover temperature and dust surface density maps. The mass in the image can be calculated with the physical pixel-size and the assumed dust-to-gas ratio $f_{dg}=\frac{\textup{dust}}{\textup{gas}}$
\begin{eqnarray}
    \label{C5:Eq:Mgas}
    M_{\textup{gas}}&=& \Sigma_{\rho} (D \vartheta_{\textup{rad}})^2 f_{dg}^{-1},
\end{eqnarray}
where $D$ is the measured distance and $\vartheta_{\textup{rad}}$ the pixel resolution in radians per pixel. 

\begin{figure*}[t]
    \includegraphics[width=\textwidth]{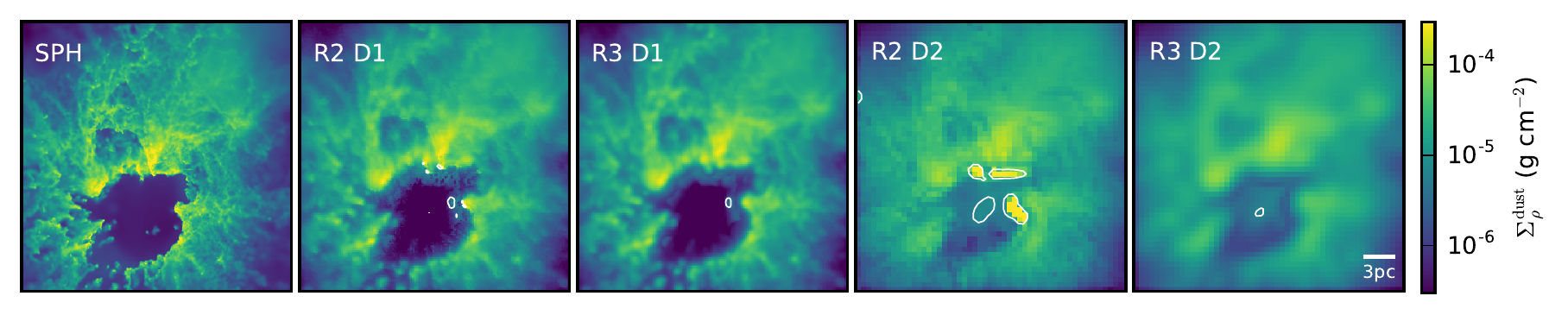}
  \caption{\label{C5:Fig:contour}Dust surface density maps from left to right: intrinsic map from the \acs{SPH} simulation, measurement for \SI{3}{\kpc} distance (\acs{D1}) and resolution version \acs{R2} and \acs{R3}, measurement for \SI{10}{\kpc} distance (\acs{D2}) and resolution version \acs{R2} and \acs{R3}. The white contours show the $\chi^2\geq5$-contour.}
\end{figure*}
%
\section{Methods | Deriving the Dust \& Gas Properties}
\label{C5:Sec:Mgas_method}
%
Before we can derive the gas mass $M_{\textup{gas}}$, we need to fit the dust surface density $\Sigma_{\rho}^{\textup{dust}}$ and the dust temperature $T_{\textup{dust}}$ from the observations (in our case synthetic observations).

We use the 5 \textit{Herschel} bands of our set of realistic synthetic observations constructed in \citetalias{KDR1:inprep}. However, some observed star-forming clouds do not show emission in the \SI{70}{\microns} band while emitting at shorter and longer wavelengths. In our analysis, we chose to make use of the \SI{70}{\microns} flux in order to recover the \ac{SED} of the modified blackbody better, and suggest that in regions where \SI{70}{\microns} flux cannot be recovered, the fit should be performed without using \SI{70}{\microns}\footnote{Note that for extra-galactic objects with significant redshift, some Herschel bands no longer traces cold dust but actually hotter material with intrinsic emission in the \ac{MIR}.}. 

\begin{itemize}
\item \textbf{Background Correction}\\
We use our realistic synthetic observations with (\acs{B2}) and without combined background (\acs{B1}). For the realistic synthetic observations with combined background (\acs{B2}), we estimate the median background emission per pixel from 12 small patches close to the border of the image and subtract the estimated background.
\vspace*{-0.15cm}
\item \textbf{Rescaling Pixel-Size}\\
The realistic synthetic observations (resolution version \acs{R1}), like real observations, have different resolutions in the different bands (see Section~\ref{C5:Sec:recap}). Therefore, before we can further follow the description of the fitting technique from Section~\ref{C5:Sec:Mgas_MBBF}, we re-grid the realistic synthetic \textit{Herschel} images to the pixel size lowest common resolution (here: $\vartheta=\ang{;;11.5}$ of \ac{SPIRE} \SI{500}{\microns}) using the \textsc{FluxCompensator}. The resulting images are labeled to the resolution version \acs{R2}.
\vspace*{-0.15cm}
\item \textbf{Convolution to Largest Common Beam Size}\\
Note that usually observers convolve all 5 \textit{Herschel} images to the largest common beam size before rescaling the pixel size (see last bullet point). This step is introduced to ensure that all images have the same "beam" resolution, ensures that within the final beam size all bands observe the same material. In this case, we would have to convolve the \acs{R1} images with a correction \ac{PSF} different for every band to recover a smoothed image similar to the smoothing in the \ac{SPIRE} \SI{500}{\microns} image (\acs{FWHM}: \ang{;;35.2}). Using the \textsc{FluxCompensator}, we therefore convolve every \acs{R1} image with a different Gaussian-shaped correction \acs{PSF} with the following standard deviation:
\begin{equation}
    \sigma_{\textup{corr}} = \sqrt{\sigma_{\textup{SPIRE \SI{500}{\microns}}}^2 - \sigma_{\textup{band}}^2}.
\end{equation}
Afterwards the pixel size (last bullet point) needs to be adjusted. The resulting images are labeled version \acs{R3}. Note that when convolving to the largest common beam more information is lost as when just rescaling the pixel size. This is why we discuss in Section~\ref{C5:Sec:Mgas_Results_Mgas} and Section~\ref{C5:Sec:Discussion} the differences arising from our analysis when not convolving to the largest common beam size but still adjusting the pixel size to the largest common resolution (version: \acs{R2}).
\vspace*{-0.15cm}
\item \textbf{Distance \& Extinction}\\
For our analysis, we assume the "correct" distances $D$ (here \SI{3}{\kpc} and \SI{10}{\kpc}) because we are only interested in the biases introduced by the fitting technique itself. Therefore, we also use the same optical extinction (here $A_V=\num{10}$ and $A_V=\num{20}$ respectively) and the same extinction law as in \citetalias{KDR1:inprep} to deredden\footnote{By deredden we mean removing the extinction from the images. This is the reverse process which has been performed when producing the realistic synthetic observations in \citetalias{KDR1:inprep}. We implement this very cautious step, because we wanted to test the raw images and not the individual changes from the optical extinction which might differ for every cloud (but still being very small). However, in real observations the reddened image including extinction is observed, therefore the assumed extinction could theoretically be removed with an extinction law and an assumed optical extinction coefficient $A_V$. However, typically the extinction for \textit{Herschel} wavelength is small and can be neglected.} the realistic synthetic observations with the \textsc{FluxCompensator} (for more details, see \citetalias{KDR1:inprep}). In reality, uncertainties about the distance and extinction would add systematic errors to the derived surface densities and temperatures.
\vspace*{-0.15cm}
\item \textbf{Choosing Dust Properties}\\
Before we can solve Eq.~\ref{C5:Eq:BBfitting} by fitting to every pixel, we need to choose a dust opacity $\kappa_\nu$ for every band. At long wavelengths, the dust opacity $\kappa_\nu$ can be approximated by a power-law in frequency:
\begin{eqnarray}
    \label{C5:Eq:kappa}
    \kappa_\nu&\sim&\nu^{\beta}
\end{eqnarray}
where $\beta$ is the power-law exponent of the dust opacity. Usually $\beta$ is fixed to $\beta\approx \num{2}$ \citep[see e.\,g.][]{Sadavoy:2014,Groves:2012,Kelly:2012,Konyves:2010,Andre:2010,DraineLi:1984}, since the fit from Eq.~\ref{C5:Eq:BBfitting} becomes degenerate in $T_{\textup{dust}}$ if $\beta$ is allowed to vary. However, there have been studies of the Magellanic Clouds by \cite{RomanDuval:2014}, where they showed that a variable $\beta$ and a variable dust-to-gas ratio is beneficial to explore local variations of the dust, but as described in \cite{Kelly:2012}, this has to be done with caution. 

In this paper, we will perform a different approach, because it is also possible to use pre-computed opacities of dust models rather than the fitted function (as Eq.~\ref{C5:Eq:kappa}). We choose the dust opacity $\kappa_\nu$ from the same model used to create the synthetic observations: \cite{Draine:2007} dust including \ac{PAH} molecules (see \citetalias{KDR1:inprep}). However, the \cite{Draine:2007} dust has three components $i$ with different opacities $\kappa_{\nu}(i)$ and different abundance ratios $f_i$. Therefore, we estimate an abundance-weighted average. 
\begin{eqnarray}
    \bar{\kappa}_{\nu} &=& \sum\limits_{i=1}^3{\kappa_{\nu}(i) f_i}
\end{eqnarray}
Further, we choose the constant dust-to-gas ratio of $\frac{\textup{dust}}{\textup{gas}}=\num{0.01}$, used in \citetalias{KDR1:inprep} to produce the synthetic observations.
\vspace*{-0.15cm}
\item \textbf{Fitting}\\
We apply the above described formalisms of the technique of modified blackbody fitting (Eq.~\ref{C5:Eq:BBfitting}) using a non-linear least-square fitting function in Python \citep[\texttt{scipy.optimize.curve\_fit}:][]{Scipy}. We produced results for our realistic synthetic observations both with \acs{B2} and without the combined background \acs{B1}. We also derive the corresponding gas masses $M_{\textup{gas}}$ following Eq.~\ref{C5:Eq:Mgas}.
\end{itemize}
\begin{itemize}
\vspace*{-0.3cm}
\item \textbf{Quality of Modified Blackbody Fit}\\
To quantify the quality of the measured (fitted) parameters $\Sigma_{\rho\textup{, measured}}^{\textup{dust}}$ and $T^{\textup{dust}}_{\textup{measured}}$, we evaluate the $\chi^2$-value\footnote{Note that since we
do not include the uncertainties, this is not technically a $\chi^2$ value, but for simplicity we will use this notation in
the remainder of this paper.} \citep[see][]{NumericalRecipes} in every pixel (c.\,f.~Eq.~\ref{C1:Eq:BBfitting1}):
\begin{eqnarray}
    \label{C5:Eq:chi_2}
    \chi^2 &\approx& \sum\limits_{\nu}{\left[\log_{10}\left(\Sigma_{B,\nu} \right)\right.} \\
    &&\left.- \log_{10}\left(\Sigma_{\rho\textup{, measured}}^{\textup{dust}} \bar{\kappa}_{\nu} B_\nu(T^{\textup{dust}}_{\textup{measured}}) \right) \right]^2\nonumber
\end{eqnarray}
The constructed $\chi^2$ maps can than be used to clip away unreliable pixels from the gas mass evaluation. In Figure~\ref{C5:Fig:contour} we show the intrinsic dust surface from the simulation and the measured dust surface density at \SI{3}{\kpc} and \SI{10}{\kpc} for the different resolution versions. The white contours show the $\chi^2>5$-contour. In Section~\ref{C5:Sec:Mgas_Results} we go into more detail about the $\chi^2$-analysis and the effects of the two different resolution versions \acs{R2} and \acs{R3} are discussed.
\end{itemize}

We note that for our analysis, we assume the correct distance $D$, optical extinction $A_V$ and dust-to-gas ratio $\frac{\textup{dust}}{\textup{gas}}$ and a comparable dust opacity $\bar{\kappa}_{\nu}$ is picked because we want to test the accuracy of the pixel-by-pixel modified blackbody fitting algorithm alone. In reality, each of these quantities will produce additional systematic uncertainties.

\begin{figure*}[t]
    \includegraphics[width=\textwidth]{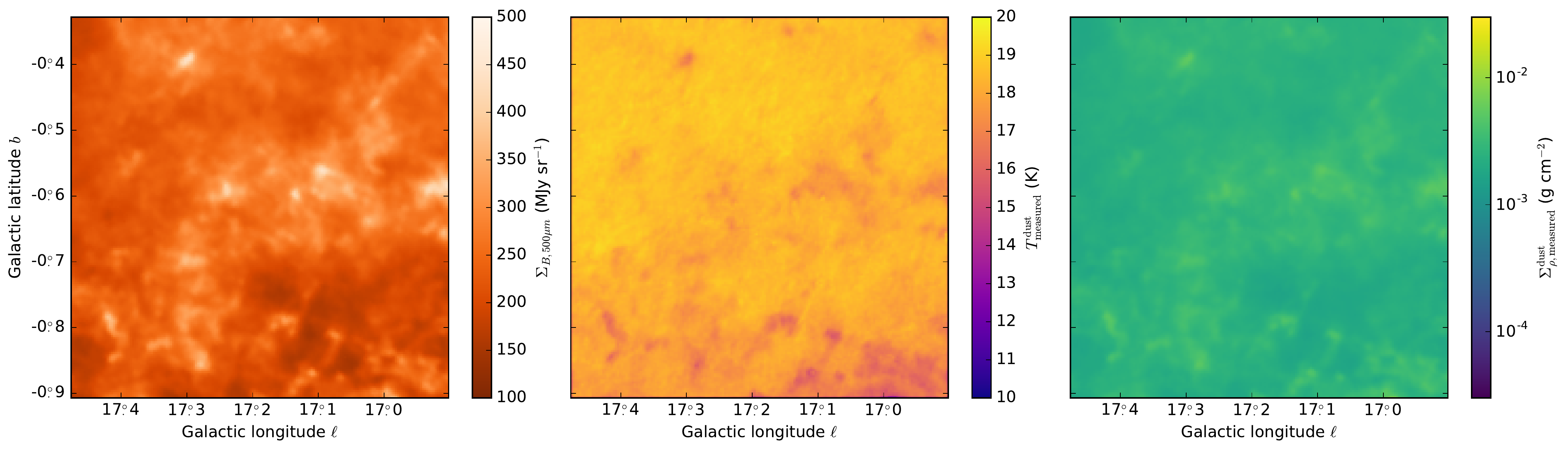}
  \caption{\label{C5:Fig:background}Observed \ac{FIR} emission, estimated dust temperature $T_{\textup{measured}}^{\textup{dust}}$ and estimated dust surface density $\Sigma_{\rho\textup{, measured}}^{\textup{dust}}$ of a relatively empty patch of the Galactic plane ($\ell=\ang{17.19128}$, $b=\ang{-0.61968587}$) with width \ang{0.575}.}
  \vspace*{+0.5cm}
\end{figure*}

\subsection{Solely Measurements of the Real Background Patch}
\label{C5:Sec:results_BB}
We used the above described method in a real-observations test-case on the 5 \textit{Herschel} observations of the empty background patch\footnote{The \textit{Herschel} images are from the \ac{Hi-GAL} and the data reduction pipeline is described in \cite{Molinari:2016}.}, which was combined with the synthetic images and resulting in the dataset \acs{B2} in \citetalias{KDR1:inprep}. We extract reasonable results for the ambient temperature in the Galactic plane. In Figure~\ref{C5:Fig:background}, we present as an example the \ac{SPIRE} \SI{500}{\microns} emission \citep[\ac{Hi-GAL} survey:][]{Molinari:2010} and the extracted dust temperature and dust surface density of the patch ($\ell=\ang{17.19128}$, $b=\ang{-0.61968587}$, width \ang{0.575}). We found a dust surface density above \SI{e-3}{\gramms\per\centi\meter\squared} and an average temperature of about \SI{18}{\kelvin}. For this reason, we used $T_{\textup{iso}}=\SI{18}{\kelvin}$ in \citetalias{KDR1:inprep} which is essential necessary to produce our realistic synthetic observations. For more details see \citetalias{KDR1:inprep}. 

\subsection{Measurements of Synthetic Images}
\label{C5:Sec:results_B2}
For the realistic synthetic observations combined with a background (\acs{B2}: for more details, see \citetalias{KDR1:inprep}), we could not perform measurements for the dust surface density or the temperature of the synthetic star-forming region because the chosen background within the Galactic plane was too high to disentangle the emission of the synthetic star-forming region (see Figure~15 in \citetalias{KDR1:inprep}). 

This is due to the fact that the \acs{D14} simulated star-forming region does not have enough mass to produce dominant emission within the Galactic plane at the longer wavelengths (from \ac{SPIRE} \SI{250}{\microns} to \ac{SPIRE} \SI{500}{\microns}; c.\,f.~with Figure~15 in \citetalias{KDR1:inprep}). However, if the synthetic observations were combined with a background of a relatively empty patch off the Galactic plane, the method would work better. 

Nevertheless, we choose a patch within the Galactic plane because most star-forming regions at a distance of \SI{3}{\kpc} lie within the Galactic plane\footnote{Closer regions and regions in the outer Galaxy (e.\,g.~Westerhout 5 also called Soul Nebula)  have lower background in \emph{Herschel} bands. If the simulations would be combined with that lower background, one could recover the gas mass. However, the  \ac{Hi-GAL}  surveys exclusively observed regions of emission (e.\,g.~star-forming regions) and never relatively empty regions on the sky off the plane.}. Therefore, for high-mass star-forming regions this method will produce measurements, while regions of lower mass will be missed within the Galactic plane. 

In the next section, we discuss the findings for the measurements without combined background (\acs{B1}).
%
\section{Results}
\label{C5:Sec:Mgas_Results}
%
In what follows we will discuss the individual results separately for the different synthetic observation setups without realistic background (\acs{B1}). We will present the measured dust surface density maps $\Sigma_{\rho\textup{, measured}}^{\textup{dust}}$, the measured dust temperature maps $T_{\textup{measured}}^{\textup{dust}}$ from modified blackbody fitting following the description of the technique from Section~\ref{C5:Sec:Mgas_method} for all our \acs{B1} realistic synthetic observations.

\subsection{Dust Surface Density and Dust Temperature Measurements}
\label{C5:Sec:Mgas_Results_SigmaTemp}
We measured the dust surface density, the dust temperature and the $\chi^2$ maps of all synthetic observations without combined background using pixel-by-pixel modified blackbody fitting. In Figure~\ref{C5:Fig:surface_density}, we present the measured dust surface density maps and in Figure~\ref{C5:Fig:temperature}, the measured dust temperatures for one orientation at a distance of \SI{3}{\kpc}. In Appendix~\ref{C5:Appendix}, we provide the maps as \acs{FITS} files for every time-step, orientation, distance and set-up version. 

In the following we discuss the results and shortcomings when using the modified blackbody fitting technique on the example time-step 122:

\begin{figure*}[p]
    \centering
    \includegraphics[width=0.85\textwidth]{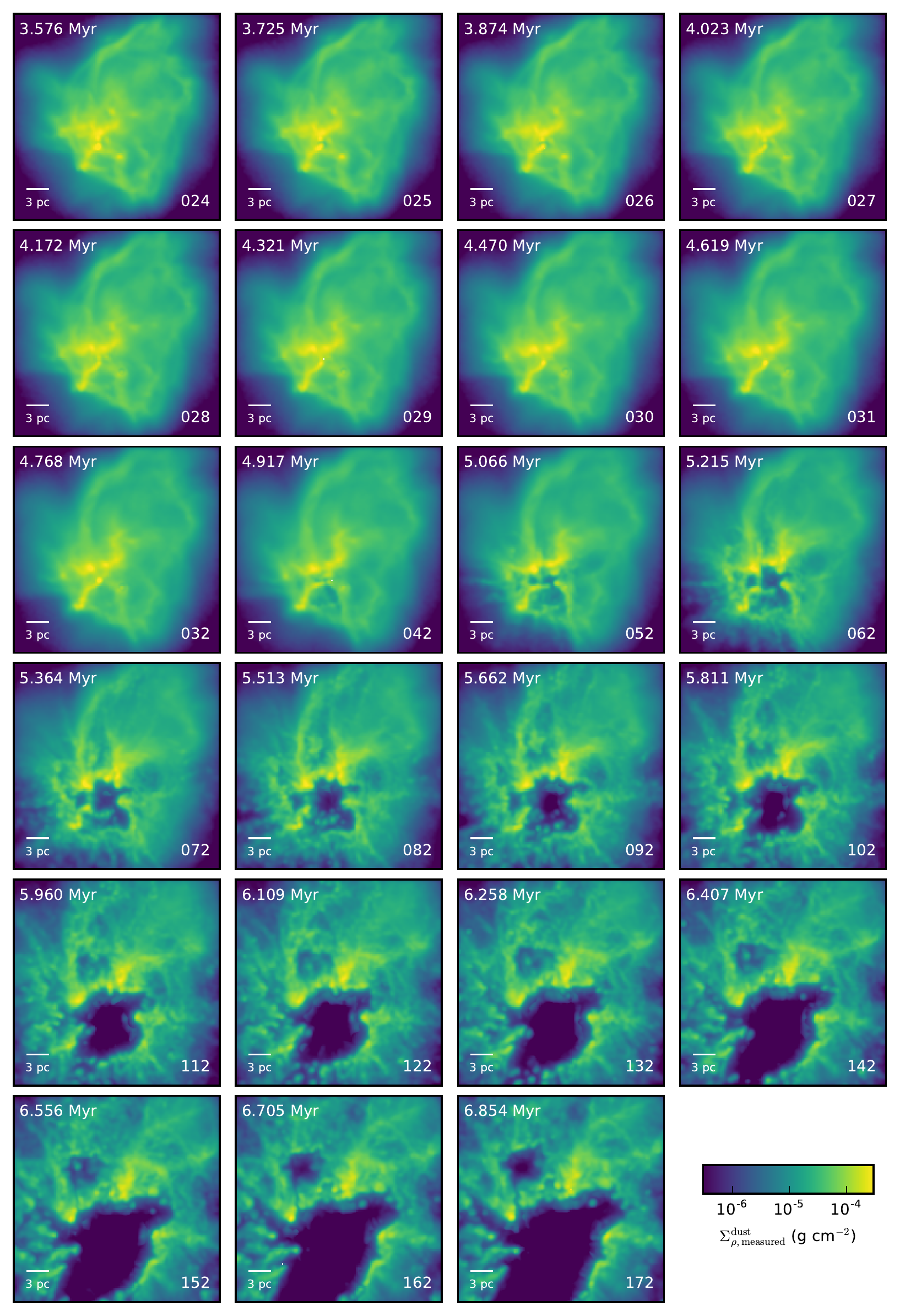}
  \caption[Measured evolution of the dust surface density]{\label{C5:Fig:surface_density}Measured dust surface density $\Sigma_{\rho\textup{, measured}}^{\textup{dust}}$ for every time-step at a distance of \SI{3}{\kpc}. The lower numbers represent the time-step IDs from the \ac{SPH} simulations. Displayed is the result for version \acs{R3}, version \acs{R2} will be available as online figure.}
\end{figure*}
\begin{figure*}[p]
    \centering
    \includegraphics[width=0.85\textwidth]{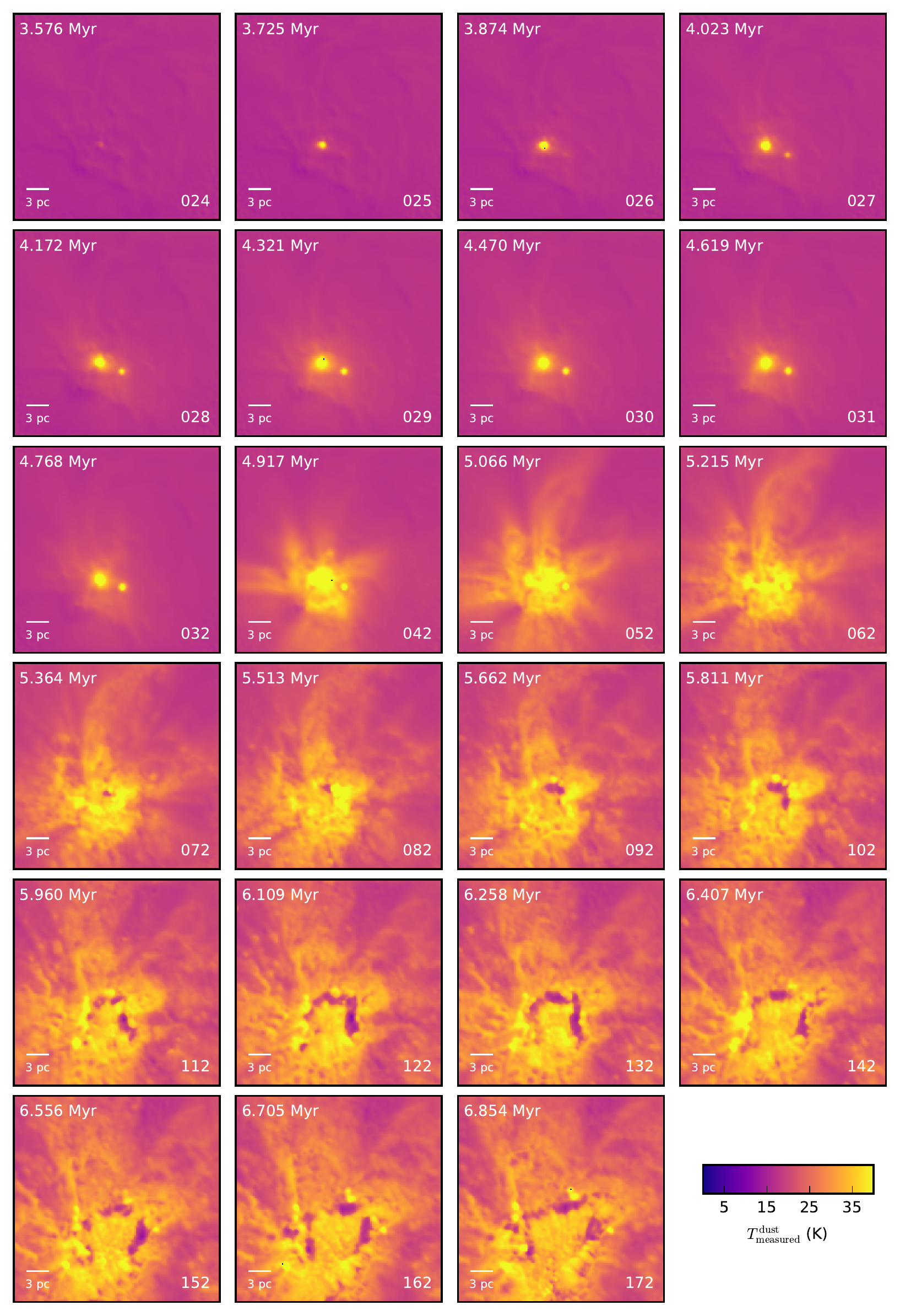}
  \caption[Measured evolution of the dust temperature]{\label{C5:Fig:temperature}Measured dust temperature $T_{\textup{measured}}^{\textup{dust}}$ for every time-step at a distance of \SI{3}{\kpc}. The lower numbers represent the time-step IDs from the \ac{SPH} simulations. Displayed is the result for version \acs{R3}, version \acs{R2} will be available as online figure.}
\end{figure*}
\begin{figure*}[t]
    \includegraphics[width=\textwidth]{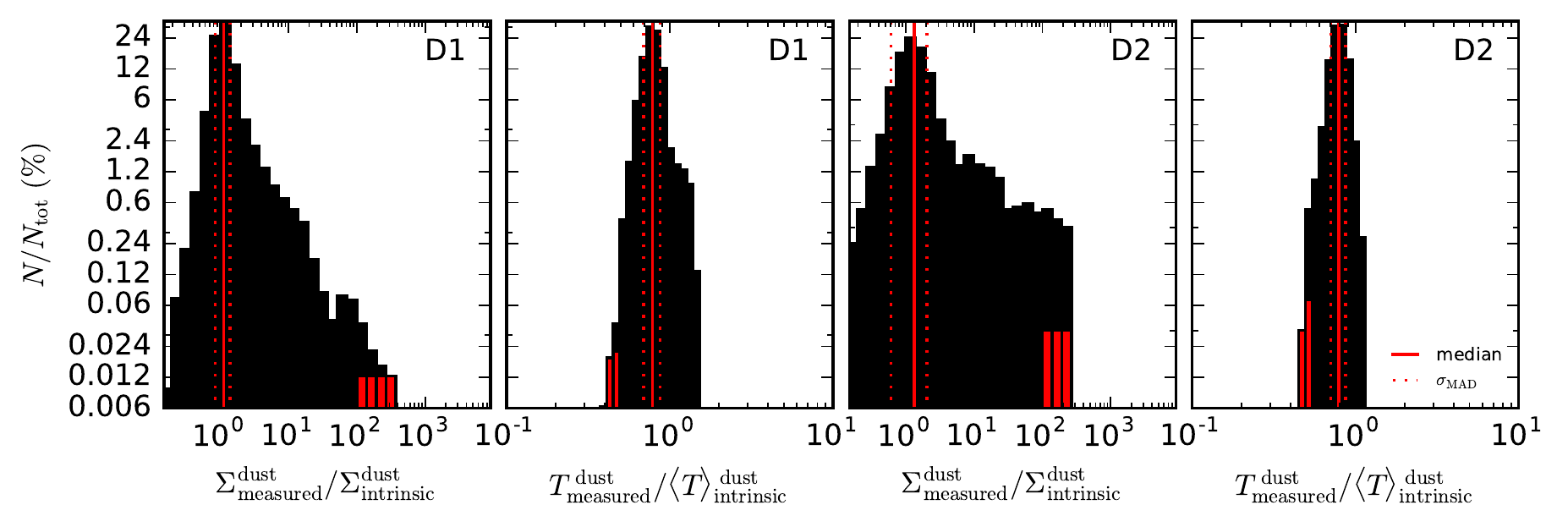}
\vspace*{-0.5cm}
  \caption{\label{C5:Fig:histogram}Log-spaced histograms of the recovered property fraction of measured value to intrinsic value (black) and respective fractions where the estimated $\chi^2>5$ (red). The red solid lines represent the median and the dashed lines the spread of the standard deviation $\sigma_{\textup{MAD}}$ estimated from the \ac{MAD}. Displayed is the result for version \acs{R3}, version \acs{R2} will be available as online figure.}
\end{figure*}
    \begin{figure*}[p]
        \centering
        \includegraphics[width=0.9\textwidth]{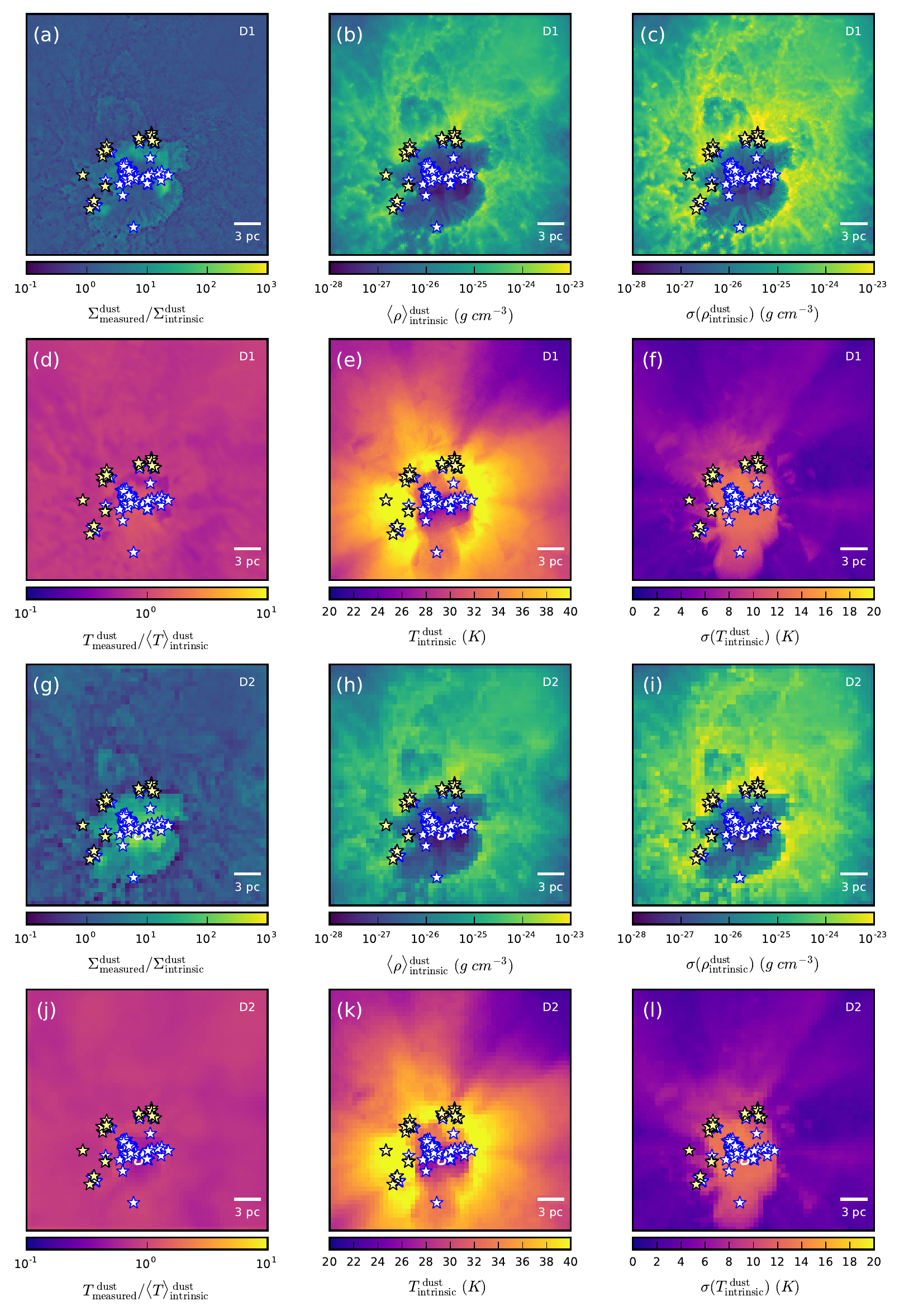}
      \caption{\label{C5:Fig:maps}Diagnostic maps of measured and intrinsic values for dust surface density, dust mean densities and dust temperatures. Upper two rows for distance \acs{D1} at \SI{3}{\kpc} and lower two rows for distance \acs{D2} at \SI{10}{\kpc}. Yellow stars are the projected positions of young, accreting stars while blue stars are stars inside the ionized bubble. $\chi^2>5$-contours are plotted in white. Displayed is the result for version \acs{R3}, version \acs{R2} will be available as online figure.}
    \end{figure*}

\subsubsection{Fitting Technique Breaks Down}
As can be seen from Figure~\ref{C5:Fig:surface_density}, the modified blackbody fitting breaks\footnote{When the fitting algorithm breaks down, the routine returns unphysical results such as negative values for $\Sigma_{\rho\textup{, measured}}^{\textup{dust}}$ and $T_{\textup{measured}}^{\textup{dust}}$.} down in very few pixels (white), which is due to the very high temperatures close to stellar sources. In these cases, the peak of the \ac{SED} for those pixels is no longer dominated by a single dust temperature and the fitting algorithm in the \textit{Herschel} bands produces unphysical 
results for $\Sigma_{\rho\textup{, measured}}^{\textup{dust}}$ and $T_{\textup{measured}}^{\textup{dust}}$. However, this only happens in a few pixels while in the majority of pixels Eq.~\ref{C5:Eq:BBfitting} produces a good fit. However, as it turns out a good fit (with physical values and moderate $\chi^2$-values) for a certain pixel does not necessarily produce a reliable result in dust temperature and dust surface density as we will show in the next sections.
    
\subsubsection{Comparing Mechanism}
    In order to compare for every pixel the intrinsic value of dust surface density or dust temperature we need to produce intrinsic projection maps of the density and temperature from the Voronoi mesh used in the radiative transfer calculation. We cannot use the temperatures and densities from the \acs{SPH} simulations directly as we refined the density structure before the radiative transfer calculation and further calculated the dust heating from the stars with the radiative transfer code. We recover the intrinsic properties from the Voronoi mesh through a Monte-Carlo approach by weighting the values of random sample points distributed along the column of every pixel. We recover the following intrinsic properties through this technique:

    \begin{itemize}
        \item intrinsic mean dust density $\langle\rho\rangle_{\textup{intrinsic}}^\textup{dust}$
        \item intrinsic standard deviation of the dust density $\sigma(\rho_{\textup{intrinsic}}^\textup{dust})$
        \item intrinsic dust surface density $\Sigma_{\textup{intrinsic}}^\textup{dust}$
        \item intrinsic mean dust temperature $\langle T\rangle_{\textup{intrinsic}}^\textup{dust}$
        \item intrinsic standard deviation of the dust temperature $\sigma(T_{\textup{intrinsic}}^\textup{dust})$
    \end{itemize}

\subsubsection{Comparison with Intrinsic Values}
\label{C5:Sec:comp}
In Figure~\ref{C5:Fig:histogram}, we show the log-scaled histograms of the fractions of measured vs. intrinsic values $\Sigma_{\rho\textup{, measured}}^{\textup{dust}} / \Sigma_{\rho\textup{, intrinsic}}^{\textup{dust}}$ and $T_{\textup{measured}}^{\textup{dust}} / \langle T\rangle_{\textup{intrinsic}}^{\textup{dust}}$ in dust surface density and in dust temperature respectively for the two different distances evaluated for every pixel of one time-step. The y-axis of the histogram gives the percentage of total images pixels in a bin. In Figure~\ref{C5:Fig:histogram}, we see that the dust surface density is overestimated in some of the pixels up to almost three orders of magnitude for close distances and the error increases when going to larger distances. When inspecting the dust temperatures, however, more pixels have measured values under-predicting the intrinsic values. We estimate the median values (MED) of the distributions for the surface density fraction and the temperature fraction and estimate the standard deviation from the median by the \acf{MAD} and the corresponding standard deviation $\sigma_{\textup{MAD}}$ using Astropy \citep{Astropy:2013}. 

We recover the following values in the format $\mbox{MED}\pm\sigma_{\textup{MAD}}$ for the example time-step 122:
    \begin{eqnarray}
        \label{C5:Eq:MED_D1_rho}
        \frac{\Sigma_{\rho \textup{, measured}}^{\textup{dust}}}{\Sigma_{\rho \textup{, intrinsic}}^{\textup{dust}}}\Bigr|_{D1}^{R3}&=&\num{1.12\pm0.28}\\
        \label{C5:Eq:MED_D1_temp}
        \frac{T_{\textup{measured}}^{\textup{dust}}}{\langle T\rangle_{\textup{intrinsic}}^{\textup{dust}}}\Bigr|_{D1}^{R3}&=&\num{0.78\pm0.09}\\
        \label{C5:Eq:MED_D2_rho}
        \frac{\Sigma_{\rho \textup{, measured}}^{\textup{dust}}}{\Sigma_{\rho \textup{, intrinsic}}^{\textup{dust}}}\Bigr|_{D2}^{R3}&=&\num{1.32\pm0.72}\\
        \label{C5:Eq:MED_D2_temp}
        \frac{T_{\textup{measured}}^{\textup{dust}}}{\langle T\rangle_{\textup{intrinsic}}^{\textup{dust}}}\Bigr|_{D2}^{R3}&=&\num{0.79\pm0.08}
    \end{eqnarray}
In relative terms this means that the translation to measured dust surface density $\Sigma_{\rho \textup{, measured}}^{\textup{dust}}$ can vary for a certain pixel from \SIrange{84}{140}{\percent} from the intrinsic value $\Sigma_{\rho \textup{, intrinsic}}^{\textup{dust}}$ for distance \acs{D1} and \SIrange{60}{204}{\percent} for distance \acs{D2}. Similar, the measured dust temperature $T_{\textup{measured}}^{\textup{dust}}$ can deviate for a certain pixel from \SIrange{69}{87}{\percent} from the real value $\langle T\rangle_{\textup{intrinsic}}^{\textup{dust}}$ for distance \acs{D1} and \SIrange{71}{87}{\percent} for distance \acs{D2}. We can see from the red bins in Figure~\ref{C5:Fig:histogram} that the pixels with high $\chi^2$-values over-predict $\Sigma_{\rho\textup{, measured}}^{\textup{dust}}$ and under-predict $T_{\rho\textup{, measured}}^{\textup{dust}}$ the most. However, removing the pixels with higher $\chi^2$-values does not affect - by definition - the median very much.

\subsubsection{Errors in Different Parts of the Cloud}
\label{C5:Sec:error}
On average (over the entire image), we recover the overall order of dust surface density and dust temperature (see Figure~\ref{C5:Fig:contour} or Figure~\ref{C5:Fig:temperature}), but the median error spread may be as high as a factor of 2 to 3 (see Section~\ref{C5:Sec:comp}) and the absolute errors of individual pixels may be exceed 2 orders of magnitude in some cases 
(see Figure~\ref{C5:Fig:histogram}). We recover the ambient dust temperature of $T_{\textup{iso}}=\SI{18}{\kelvin}$, which we put in the radiative transfer calculation (see \citetalias{KDR1:inprep}) and we can observe the additional heating due to the stellar radiation. In the following, we inspect how the different properties change as a function of position in the cloud. We produced 2d-maps of the intrinsic properties and intrinsic-to-measured property ratios in Figure~\ref{C5:Fig:maps} for distance \acs{D1} and \acs{D2}. We found that the measured vs. intrinsic surface density ratios (Figure~\ref{C5:Fig:maps}, a, g) is highest in low-density regions with $\langle\rho\rangle_{\textup{intrinsic}}^\textup{dust}<\SI{1e-26}{\gccm}$ (see Figure~\ref{C5:Fig:maps}, b, h) or $\Sigma_{\textup{intrinsic}}^\textup{dust}<\SI{1e-6}{\gcm}$. Further, we found that errors arise in pixels which have not only low densities but also high temperature dispersion along the line of sight (Figure~\ref{C5:Fig:maps}, f, l), and which are generally cooler than the surrounding (Figure~\ref{C5:Fig:maps}, e, k). We also note that the error is largest for regions close to the rim of the ionized bubble\footnote{The cavity of the ionized bubble is opened by high-mass stellar feedback in the center of the cloud after time-step \num{032} and can be considered as the H$_{II}$ region of the star-forming region.}, where there is a spatial density gradient and young accreting stars near-by (Figure~\ref{C5:Fig:maps}, b, h). The radiation from the stars causes the large temperature dispersion along the line of sight which is also favored by a gradient in density both along the line of sight and spatially (Figure~\ref{C5:Fig:maps}, b, c, h, i). The errors increase for larger distances where the physical size of a pixel gets larger and with it the gradient. Generally, we see (Figure~\ref{C5:Fig:maps}, a, d, g, j and Figure~\ref{C5:Fig:histogram}) that an overestimation of the dust surface density leads to an underestimation of the dust temperature from the modified blackbody fitting technique.

    \begin{figure}[t]
        \includegraphics[width=\textwidth]{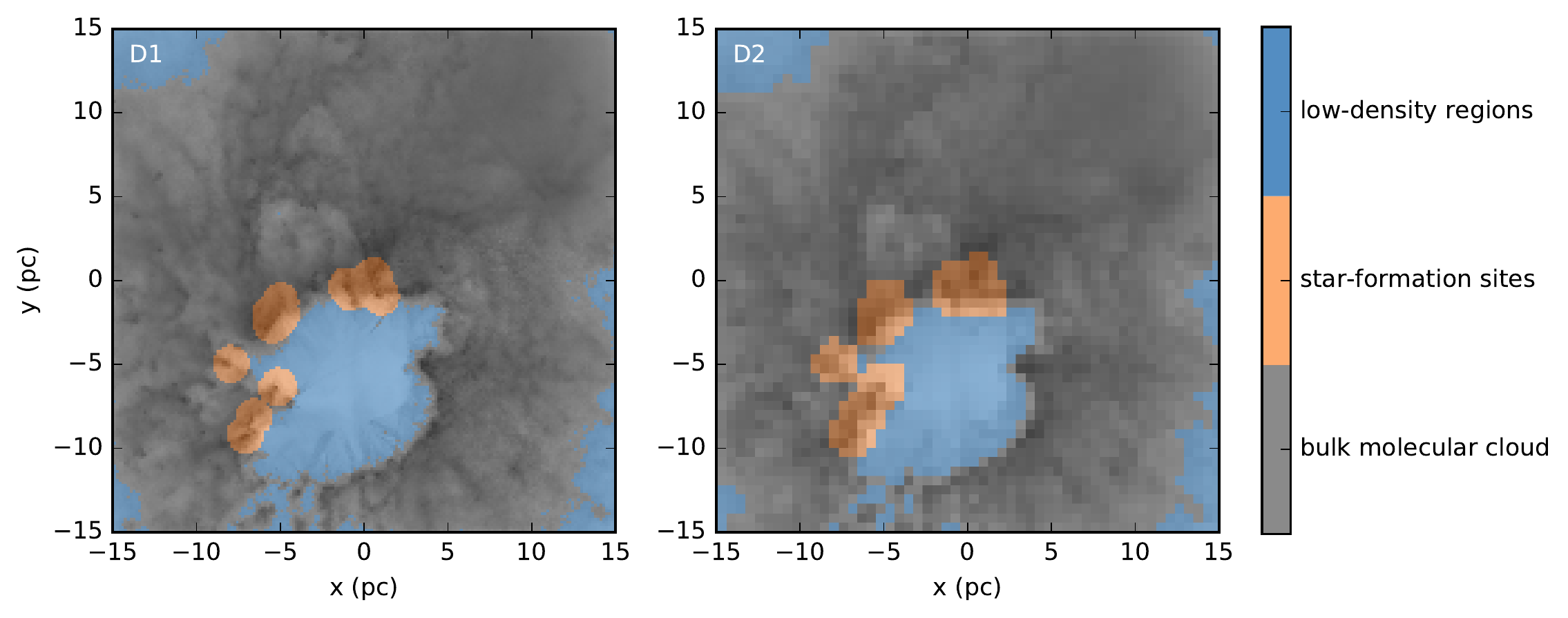}
        \vspace*{-0.5cm}
      \caption{\label{C5:Fig:regions_122}Morphology of the intrinsic column density maps overlain with the selected subregions following our criteria from Eq.~\ref{C5:Eq:criteriaI} to Eq.~\ref{C5:Eq:criteriaIII}. Left for distance \acs{D1} at \SI{3}{\kpc} and right for distance \acs{D2} at \SI{10}{\kpc}.}
    \end{figure}

\begin{figure*}[p]
    \centering
    \includegraphics[width=0.9\textwidth]{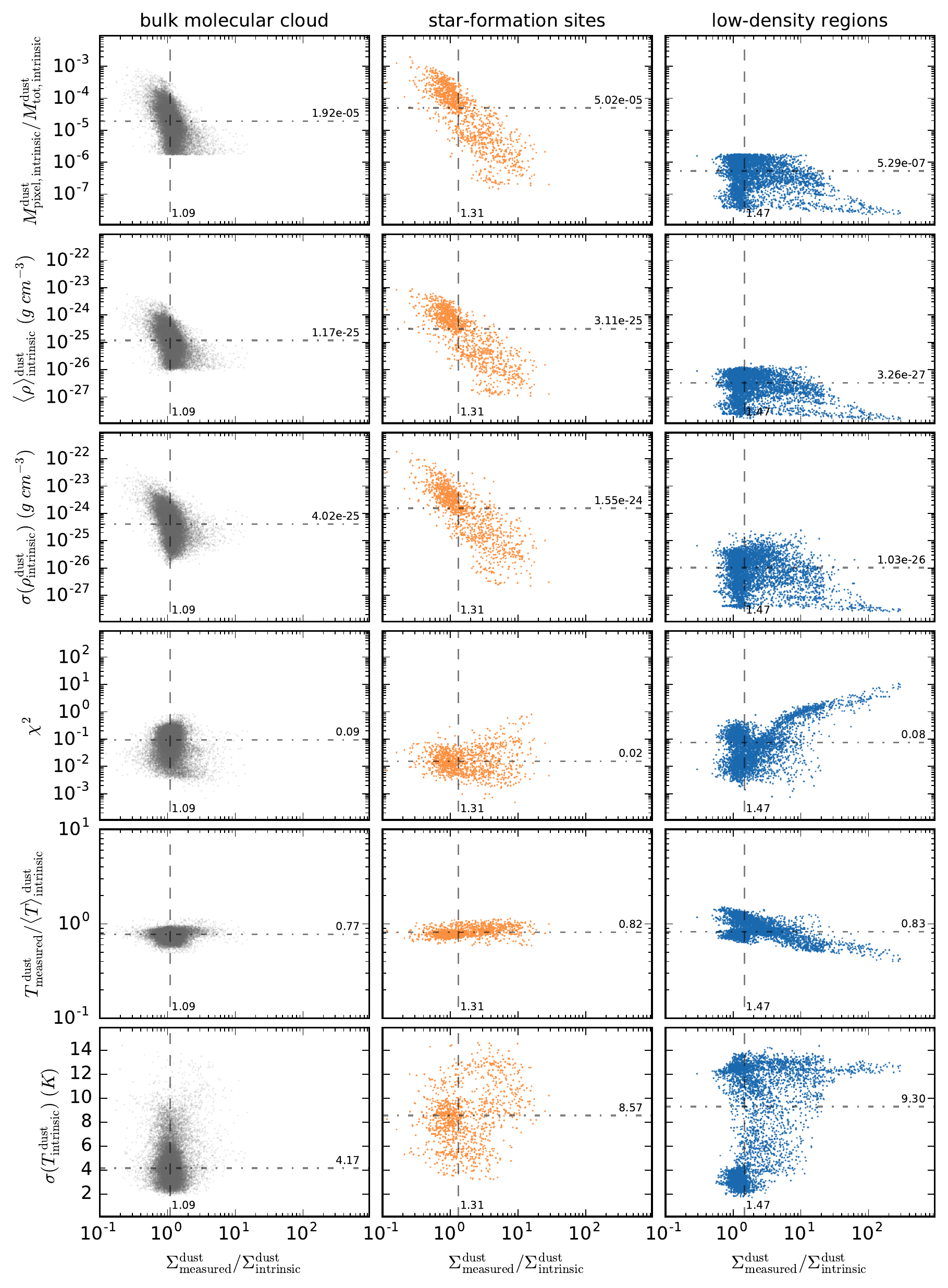}
  \caption{\label{C5:Fig:scatter_D1}Pixel-by-pixel analysis of the intrinsic properties and measured properties in dust surface density, dust density and dust temperature for distance \acs{D1} at \SI{3}{\kpc}. The dashed black lines mark the median value in x and y with the corresponding values. Displayed is the result for version \acs{R3}, version \acs{R2} will be available as online figure.}
\end{figure*}
\begin{figure*}[p]
    \centering
    \includegraphics[width=0.9\textwidth]{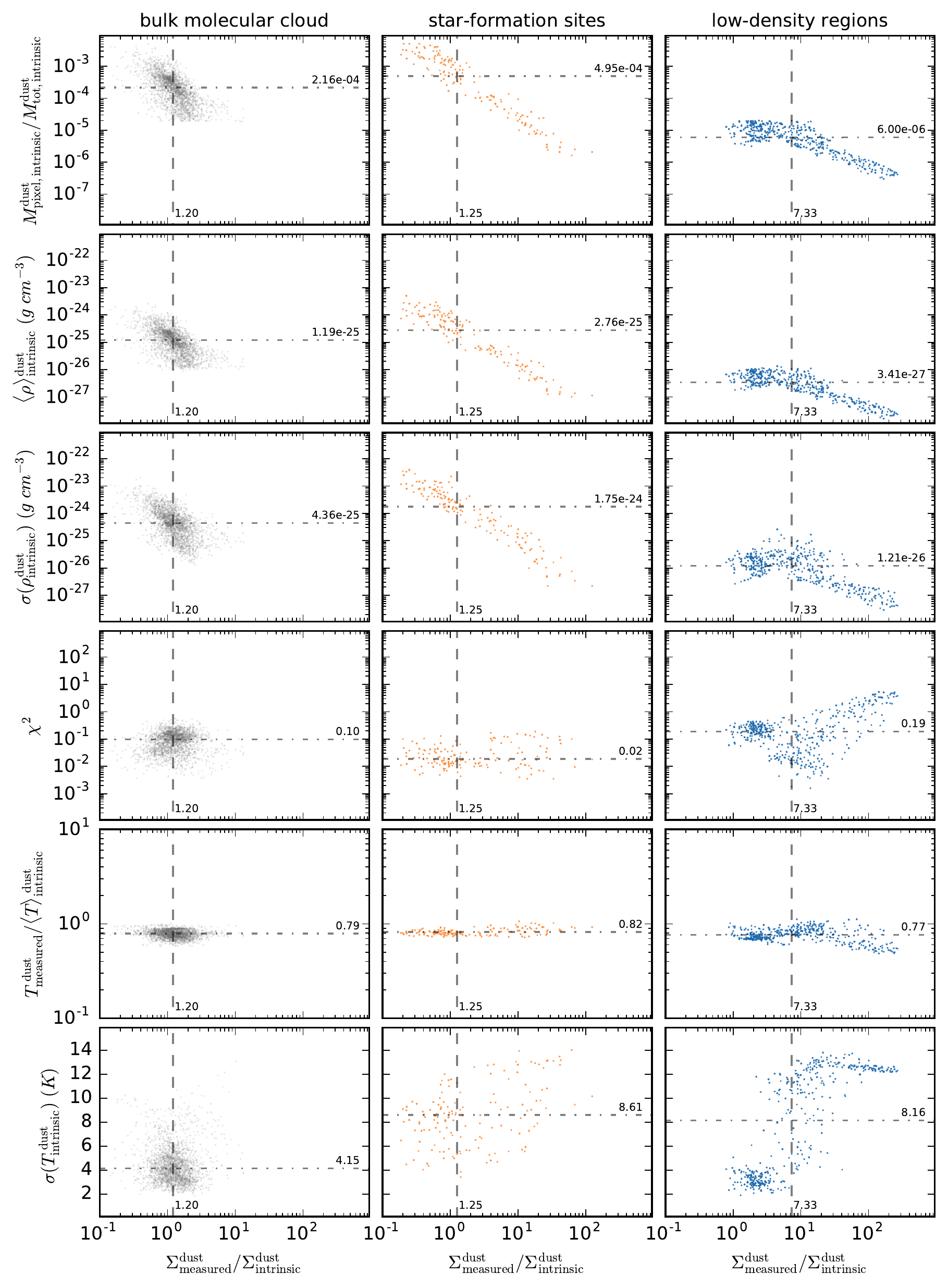}
  \caption{\label{C5:Fig:scatter_D2}Pixel-by-pixel analysis of the intrinsic properties and measured properties in dust surface density, dust density and dust temperature for distance \acs{D2} at \SI{10}{\kpc}. The dashed black lines mark the median value in x and y with the corresponding values. Displayed is the result for version \acs{R3}, version \acs{R2} will be available as online figure.}
\end{figure*}

\begin{figure*}[p]
    \centering
    \includegraphics[width=0.9\textwidth]{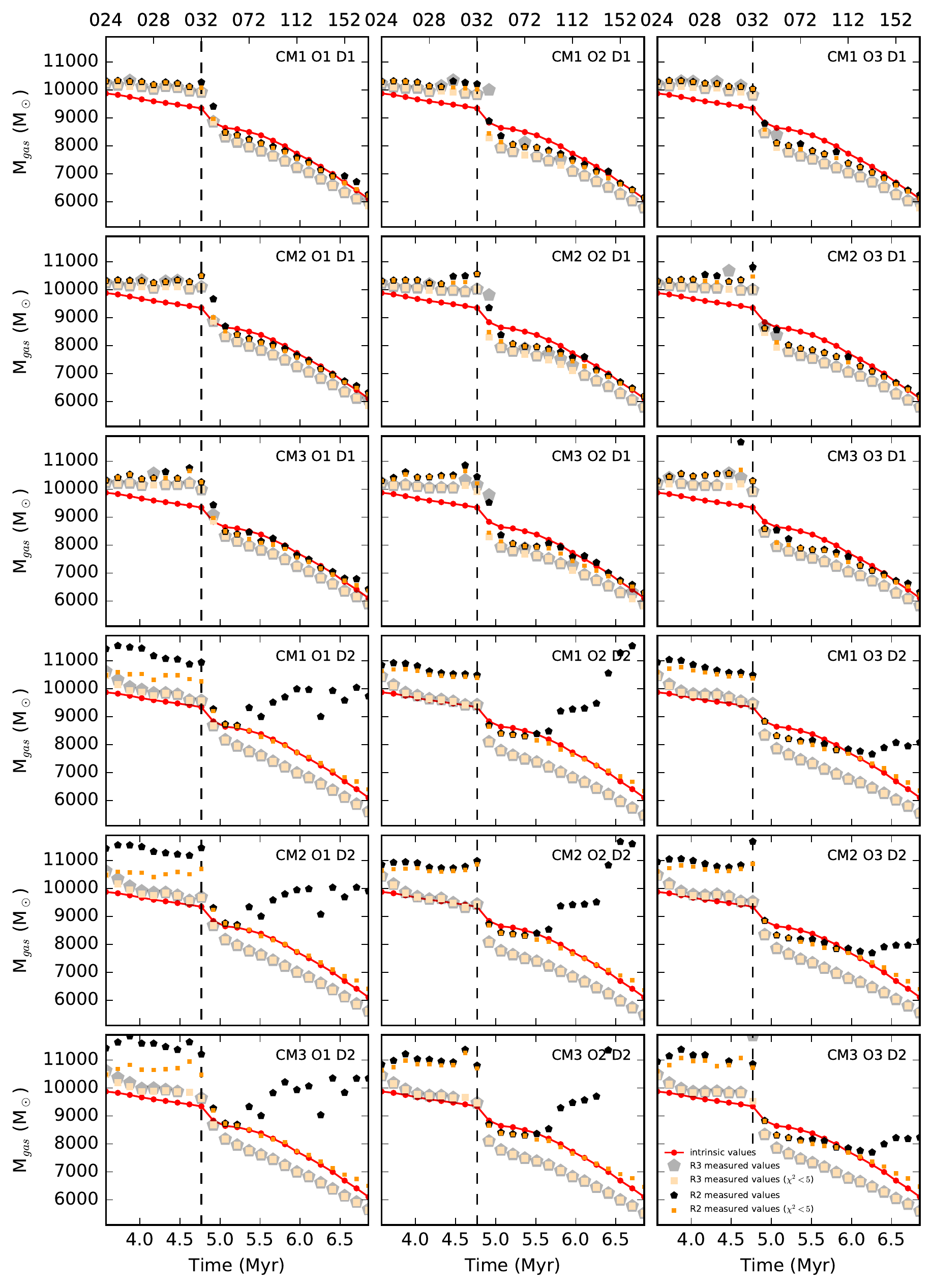}
  \caption{\label{C5:Fig:Mgas}Measured gas masses with modified blackbody fitting for the different orientations (\acs{O1}: xy plane, \acs{O2}: xz plane, \acs{O3}: yz plane), circumstellar set-ups (\acs{CM1}: no refinement, \acs{CM2}: refinement with rotationally flattened envelope, \acs{CM3}: refinement with power-law envelope) and distances (\acs{D1}: \SI{3}{\kpc}, \acs{D2}: \SI{10}{\kpc}). The top numbers represent the time-step IDs from the \ac{SPH} simulations. Vertical dashed lines mark the onset of high-mass stellar feedback.}
\end{figure*}

\subsubsection{Pixel-by-Pixel Analysis of Sub-regions}
Following the last paragraph, we divide the star-forming cloud up in 3 sub-regions (see Figure~\ref{C5:Fig:regions_122}):

    \begin{itemize}
        \item sub-region I: bulk molecular cloud\footnote{The cloud without the sub-region I and the sub-region II.}
            \begin{equation}
                \label{C5:Eq:criteriaI}
                \ \ \ \ \ \ \ \ \ \ \mbox{cloud} \setminus (\mbox{sub-region II}\ \cup\ \mbox{sub-region III})
            \end{equation}
        \item sub-region II: star-formation sites\\
            nearest-neighbor pixels to projected positions of accreting stars with maximal distance close to \num{10}-times the adopted outer envelope radius from \citetalias{KDR1:inprep}:
            \begin{equation}
                d_{max}=\SI{1}{\pc} + \frac{\sqrt{2}}{2} \mbox{pixel size}
            \end{equation} 
        \item sub-region III: low-density regions
            \begin{eqnarray}
                \label{C5:Eq:criteriaIII}
     \langle\rho\rangle_{\textup{intrinsic}}^\textup{dust}&<&\SI{e-26}{\gccm}\\
            \mbox{or}\ \ \ \ \  \Sigma_{\textup{intrinsic}}^\textup{dust}&<&\SI{e-6}{\gcm}\nonumber
            \end{eqnarray}
    \end{itemize}
In Figure~\ref{C5:Fig:regions_122}, we visualize the 3 sub-regions estimated from the criteria in Eq.~\ref{C5:Eq:criteriaI} to Eq.~\ref{C5:Eq:criteriaIII} for the two different distances. The sub-regions differ a little for the different distances since the pixel-size varies as well. We created scatter plots (\acs{D1}: Figure~\ref{C5:Fig:scatter_D1}, \acs{D2}: Figure~\ref{C5:Fig:scatter_D2}) of the pixel values for combinations of intrinsic star-formation properties and measured star-formation properties similar to the 2d-maps in Figure~\ref{C5:Fig:maps}. We highlight the median values of the distributions with vertical and horizontal lines. In the following, we discuss the findings when inspecting these plots: From the x-axis values of all the panels in Figure~\ref{C5:Fig:scatter_D1} (\acs{D1}) and Figure~\ref{C5:Fig:scatter_D2} (\acs{D2}) we can see that the fraction of $\Sigma_{\rho\textup{, measured}}^{\textup{dust}} / \Sigma_{\rho\textup{, intrinsic}}^{\textup{dust}}$ is close to unity and to the total cloud value (c.\,f.~Eq.~\ref{C5:Eq:MED_D1_rho}) for the bulk mass of the cloud, while the median for the star-forming sites and the low-density regions is further\\

\begin{minipage}{\textwidth}
\vspace{+7cm}
\end{minipage}

away from unity. The median values move further away from the total cloud value (c.\,f.~Eq.~\ref{C5:Eq:MED_D2_rho}) and unity the larger the distance gets, due to the effects described in Section~\ref{C5:Sec:error}. At distance \acs{D2}, for low-density regions the median of the fraction is \num{7.33}, which means that the measured value exceeds more than \num{7}-times the intrinsic value. We list the median fraction of all the sub-regions and time-steps, orientation and distances in Table~\ref{C5:Appendix:median} in Appendix~\ref{C5:Appendix} including the corresponding $\sigma_{\textup{MAD}}$ and resulting relative errors. The contributing dust mass fraction $M_{\textup{pixel}}^{\textup{intrinsic}} / M_{\textup{tot}}^{\textup{intrinsic}}$ (first row of  Figure~\ref{C5:Fig:scatter_D1}) is highest for the star-forming sub-region (orange) which is not surprising as the stars are formed in very dense environments. The contribution of the low-density regions is lowest (by definition). Therefore, large fitting errors in the ionized bubble do not contaminate the total gas mass (see Section~\ref{C5:Sec:Mgas_Results_Mgas}) too much. When comparing Figure~\ref{C5:Fig:scatter_D1} (\acs{D1}) with Figure~\ref{C5:Fig:scatter_D2} (\acs{D2}), we can see that the median values are larger - this is due to the increase in the physical size of the pixels. In the second and third row of the scatter plots we show the intrinsic mean dust density $\langle\rho\rangle^{\textup{dust}}_{\textup{intrinsic}}$ and the corresponding standard deviation $\sigma(\rho^{\textup{dust}}_{\textup{intrinsic}})$. We note that the order of magnitude of the intrinsic mean dust density $\langle\rho\rangle^{\textup{dust}}_{\textup{intrinsic}}$ is comparable to the order of magnitude of the corresponding standard deviation $\sigma(\rho^{\textup{dust}}_{\textup{intrinsic}})$. Star-forming sites have higher mean densities than other parts of the cloud and a larger spread in dust density standard deviation. This is expected as compact clumps that form stars can have densities that can vary over several orders of magnitude from the surrounding cloud (see Figure~8 in \citetalias{KDR1:inprep}). In the fourth row, we plot the estimated $\chi^2$-values for every pixel. Most $\chi^2$-values show very low $\chi^2$-values hinting at good fits of the measured \acp{SED}. Further, high $\chi^2$-values follow the largest surface density over-estimate in the low-density regions (see also Figure~\ref{C5:Fig:histogram}). The measured vs. intrinsic fraction of the dust temperature $T_{\textup{measured}}^{\textup{dust}} / \langle T\rangle_{\textup{intrinsic}}^{\textup{dust}}$ is plotted in the fifth panel. We can see that the median of the fraction is less than unity. This underestimation grows by \SI{6}{\percent} for low-density regions at larger distances (see Figure~\ref{C5:Fig:scatter_D2}) coinciding with regions of density overestimation. We list the median fraction of all the sub-regions and time-steps, orientation and distances in Table~\ref{C5:Appendix:median} in Appendix~\ref{C5:Appendix} including the corresponding $\sigma_{\textup{MAD}}$ and relative errors. In the last row of Figure~\ref{C5:Fig:scatter_D1}, we plot the standard deviation of the intrinsic dust temperature $\sigma(T^{\textup{dust}}_{\textup{intrinsic}})$ in the line of sight of every pixel. From Figure~\ref{C5:Fig:scatter_D1}, we can see that the median of the standard deviation is lowest for the sub-regions of the bulk molecular cloud and more than double for the other sub-regions regardless of distance. When inspecting Figure~\ref{C5:Fig:scatter_D1} and Figure~\ref{C5:Fig:scatter_D2} we find that the pixels with $\sigma(T^{\textup{dust}}_{\textup{intrinsic}}) > \SI{6}{\kelvin}$ are more likely to have overestimated dust surface densities which also coincides with an underestimation of dust temperature. This is especially true for sub-regions of the cloud which contain star-formation sites or low-density regions observed at large distances.

\begin{figure*}[t]
    \includegraphics[width=\textwidth]{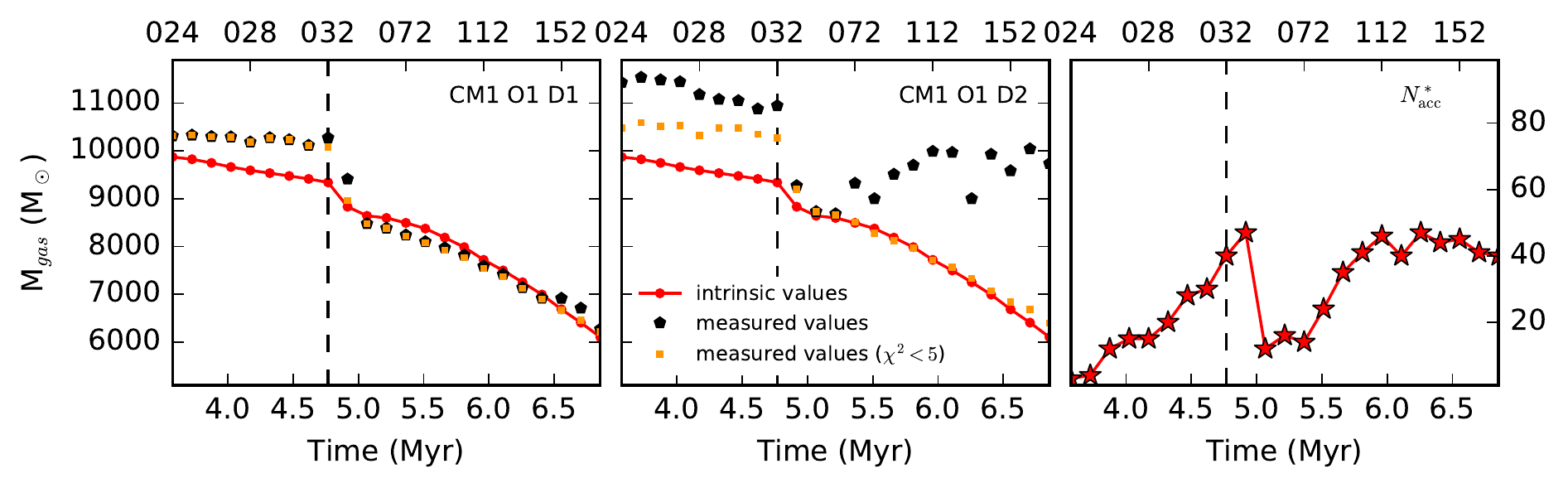}
\vspace*{-0.5cm}
  \caption{\label{C5:Fig:stars_vs_mass_D2}Total gas mass measurements (left, middle) for two representative setups extracted from surface density maps (resolution version \acs{R2}) and the evolution of the number of accreting stars $N^*_{\textup{acc}}$ (right). Vertical dashed lines mark the onset of high-mass stellar feedback.}
\end{figure*}

\subsection{Gas Mass Measurements}
\label{C5:Sec:Mgas_Results_Mgas}
For the \acs{B1} synthetic observations using Eq.~\ref{C5:Eq:Mgas} and the dust surface density maps from Section~\ref{C5:Sec:Mgas_Results_SigmaTemp}, the gas mass $M_{\textup{gas}}$ is evaluated for all the time-steps, orientations, distances and set-up versions and we provide the measured values in Table~\ref{C5:Appendix:median} in Appendix~\ref{C5:Appendix}. 

\subsubsection{$\chi^2$-Anlaysis}
\label{C5:Sec:Mgas_Results_chi}
In Figure~\ref{C5:Fig:contour} for one time-step, we see the changes in the surface density and $\chi^2$-contours (white) introduced by distance as well as the differences introduced by the different resolution approaches \acs{R2} and \acs{R3}. For the closer distance (\acs{D1}) we see that fewer pixels have values of $\chi^2>5$ in the version \acs{R3} than in \acs{R2}. At first glance this is strange, because of course the underlying temperature structure along the line-of-sight is the same. The difference is due to the convolution in version \acs{R3} to the largest common beam size. Through this convolution information from the shorter wavelength bands with smaller initial beam sizes is lost. Therefore, the fluxes from pixels which have an initial large erroneous temperature are washed out over the area of the beam, "contaminating" larger areas. Of course due to the convolution the flux also gets weaker at the originating pixel. This circumstance leads to lower $\chi^2$-quantities, which mistakenly suggest a "better" fit, although the originating pixel and the area around it (due to the \acs{R3} convolution) are corrupted. This fact also becomes apparent when comparing to larger distances. For the \acs{R3} at larger distances, $\chi^2>5$ contours move in their spatial position comparing to the closer counterpart, while for the \acs{R2} method the areas with $\chi^2>5$ just grow in size for larger distances. We also note that high $\chi^2$-values are more strongly coupled with surface density overestimation and dust temperature underestimation than for the version \acs{R3}. As an example, areas with high $\chi^2$-values in low-density regions coincide with large surface density overestimation and can be therefore used to improve the total gas mass measurements for surface density and dust temperature maps but only when extracted from \acs{R2} synthetic images.

For \acs{R2} maps, most $\chi^2$-values from the \SI{3}{\kpc} sample are very low hinting at good fits of the measured \acp{SED}. For larger distances, a considerable fraction of pixels has values of $\chi^2>5$ especially in the sub-regions of star-forming sites and low-density regions. Further, $\chi^2$-values follow the largest surface density over-estimate in the low-density regions, but are considerably higher than for method \acs{R3} displayed in Figure~\ref{C5:Fig:histogram}. This is not surprising as for the \acs{R3} method the fluxes of a pixel which would result in a bad fit have been washed out by the convolution making the $\chi^2$-value untrustworthy by conception.

To summarize, when convolving to the largest beam size more information and structure is lost (see Figure~\ref{C5:Fig:contour}). $\chi^2$-analysis can only be used consistently if no convolution to the largest common beam is performed. $\chi^2$-values maps which resulted from \acs{R3} maps are faulty by conception, since low $\chi^2$-values are not a good indicator for a good fit and the high $\chi^2$-contours indicate now enlargements for larger distances which is expected. Therefore, when no final convolution is applied (\acs{R2}), then the overestimation in certain pixels with low-density regions will grow especially for large distances. Then removing pixels with large $\chi^2$-values can help to mitigate the contamination from those pixels resulting in a more reliable result.\\

In the following we will discuss the changes on the total gas mass measurements when using surface density maps contracted with \acs{R2} and \acs{R3} synthetic images.

\subsubsection{Time-evolution}
We plot the time-evolution of the measured (black) and the intrinsic (red) gas mass in Figure~\ref{C5:Fig:Mgas}. The measurements from the \acs{R2} version are plotted small and solid while the \acs{R3} measurements appear larger and opaque. Note that the gas mass decreases over time because gas moves outside of the field of view due to the high-mass stellar feedback.

For the closer distance runs (\acs{D1}, \SI{3}{\kpc}), we can see that the measured \acs{R3} gas mass closely follows the gas mass from the simulation. There are differences of at most \SI[retain-explicit-plus]{+9}{\percent} for the time-steps without feedback before \SI{4.7}{\Myr} and \SI{-10}{\percent} for the high-mass stellar feedback phase. The \acs{R2} errors for the pre-feedback times are at most \SI[retain-explicit-plus]{+14}{\percent} for high angular resolution (\acs{D1}, \SI{3}{\kpc}) from the actual values. For the high-mass stellar feedback dominated time-steps the errors are considerably lower. They vary between \SI[retain-explicit-plus]{-7}{\percent} and \SI[retain-explicit-plus]{+4}{\percent} for distance \acs{D1}.

In Figure~\ref{C5:Fig:Mgas} we see that for close distances the pre-feedback time-steps produce a too high total gas mass. In these time-steps, two objects larger than \SI{20}{\Msun} are still accreting without driving winds or ionizing the region around. This is due to the unorthodox set-up of the \acs{D14} simulations in time-steps before the high-mass stellar feedback is switched on. These objects produce a temperature structure in their surrounding with a large dynamic range along the line-of-sight whereas modified blackbody fits assume a single temperature component, and this is (as was also shown in Section~\ref{C5:Sec:Mgas_Results_SigmaTemp}) the cause of the discrepancy. Once feedback is switched on the high-mass accreting objects ionize the region and also drive winds as they should have done before, but this was suppressed by the simulation setup of the \acs{D14} simulations. The onset of the high-mass stellar feedback removes material away from these high-mass objects and the close dust is destroyed and the objects and their immediate surroundings become invisible in the infrared. Less contamination of the line-of-sight temperature profile causes the technique to produce more reliable results as can be seen in the time-steps after the simulation switch.

\subsubsection{Distances}
For the distant run (\acs{D2}, \SI{10}{\kpc}), we see that the \acs{R3} measured gas mass deviates a little more from the actual gas mass once feedback is switched on (\SI{-13}{\percent}). In comparison (see Figure~\ref{C5:Fig:Mgas}), we see that the non-$\chi^2$-corrected \acs{R2} measured gas mass deviates more strongly from the actual gas mass, by up to a factor of \num{2}. The increase of error for larger distances is due to a larger fraction of the mass which ends up in every pixel with a wider distribution of temperatures, and which violates the single-temperature assumption of the technique. The error at larger distances gets especially large at later time-steps after about \SI{5.5}{\Myr} (see Section~\ref{C5:Sec:errorevo}).

From the analysis of the $\chi^2$-values and the corresponding maps (see Section~\ref{C5:Sec:Mgas_Results_SigmaTemp} and Section~\ref{C5:Sec:Mgas_Results_chi}), we found that for the images which where convolved to the largest common beam size (\acs{R3}) the measurements cannot be improved by $\chi^2$ analysis, since the $\chi^2$-values do not resemble the quality of the fit.

From the \acs{R2} analysis of the $\chi^2$-values (see Section~\ref{C5:Sec:Mgas_Results_SigmaTemp} and Section~\ref{C5:Sec:Mgas_Results_chi}) and the corresponding maps, we found that pixels with a $\chi^2>\num{5}$ are least reliable and especially when they occur in low-density regions. While the close-by runs (\acs{D1}, \SI{3}{\kpc}) lie mostly below $\chi^2=\num{5}$ (c.\,f.~Figure~\ref{C5:Fig:contour}), for the distant run (\acs{D2}, \SI{10}{\kpc}) the $\chi^2$-values lie above the threshold close to the ionized bubble. In these cells within the low-density region, there are large spatial and line-of-sight density and temperature gradients due to the presence of protostars (see Section~\ref{C5:Sec:Mgas_Results_SigmaTemp}). When clipping the corresponding pixels above the $\chi^2$-threshold away, we recover a gas mass which is comparable to the intrinsic value. As in Figure~\ref{C5:Fig:Mgas}, we highlighted the $\chi^2$-corrected \acs{R2} gas masses $M_{\textup{gas}}^{\chi^2<5}$ with solid orange squares. We can see that $\chi^2$-corrected \acs{R2} total gas mass measurements are not considerably underestimated once high-mass stellar feedback is switched on, in contrast to the \acs{R3} measurements displayed in Figure~\ref{C5:Fig:Mgas}, regardless of distance.

\subsubsection{Orientations \& Sources Setup}
Furthermore, from our analysis we found that the contamination is different for different orientations (\acs{O1}: xy plane, \acs{O2}: xz plane, \acs{O3}: yz plane). This is plausible, since for different orientations the stars form in different pixels and sometimes might be aligned into one and the same. These differences vary the temperature structure along the line-of-sight. In addition, the type of circumstellar set-up (\acs{CM1}: no refinement, \acs{CM2}: refinement with rotationally flattened envelope, \acs{CM3}: refinement with power-law envelope) affect the error as well, since the setup affects the distribution of the heated material close to the young stars.

\subsubsection{Error Evolution}
\label{C5:Sec:errorevo}
A combination of the above discussed biases is the reason for the increasing non-$\chi^2$-corrected \acs{R2} error after \SI{6}{\Myr} for the distant regions (\acs{D2}, middle panel of Figure~\ref{C5:Fig:stars_vs_mass_D2} or Figure~\ref{C5:Fig:Mgas}). In comparison with Table~1 of \citetalias{KDR1:inprep} and the right panel of Figure~\ref{C5:Fig:stars_vs_mass_D2} we know that more accreting objects are present at later time-steps of \textit{run I} in the \acs{D14} simulations. We note that the once feedback is switched on the gas mass measurement (black pentagons) follows the number of accreting stars $N^*_{\textup{acc}}$ (red stars). Further for later time-steps there are more accreting objects present but they are also more spatially distributed compared to the time-steps before the high-mass stellar feedback initiated. More heating stars which fall in different pixels lead to an overestimate which is especially high for larger distances.\\

Overall, the technique is successful in measuring the total gas mass for the different orientations and radiative transfer set-ups. Method \acs{R2} produced considerably lower errors especially in the high-mass stellar feedback phase. In Section~\ref{C5:Sec:Discussion_distances} we will give error limits for the different techniques.

Whether the technique still works for more complex systems, such as distant galaxies, where a pixel combines several \si{pc}, needs to be tested in future. Additionally, we have to keep in mind that once a realistic background is included, the errors will grow, but these errors will then be due to overestimation or underestimation of the non-uniform background (see Section~\ref{C5:Sec:results_BB}).

\section{Discussion}
\label{C5:Sec:Discussion}
In the sections above, we tested a technique commonly applied by observers to measure star-formation properties such as the dust surface density, the dust temperature and the gas mass. We will now discuss the shortcomings and challenges when testing and measuring these properties:

\begin{figure*}[p]
    \centering
    \includegraphics[width=0.85\textwidth]{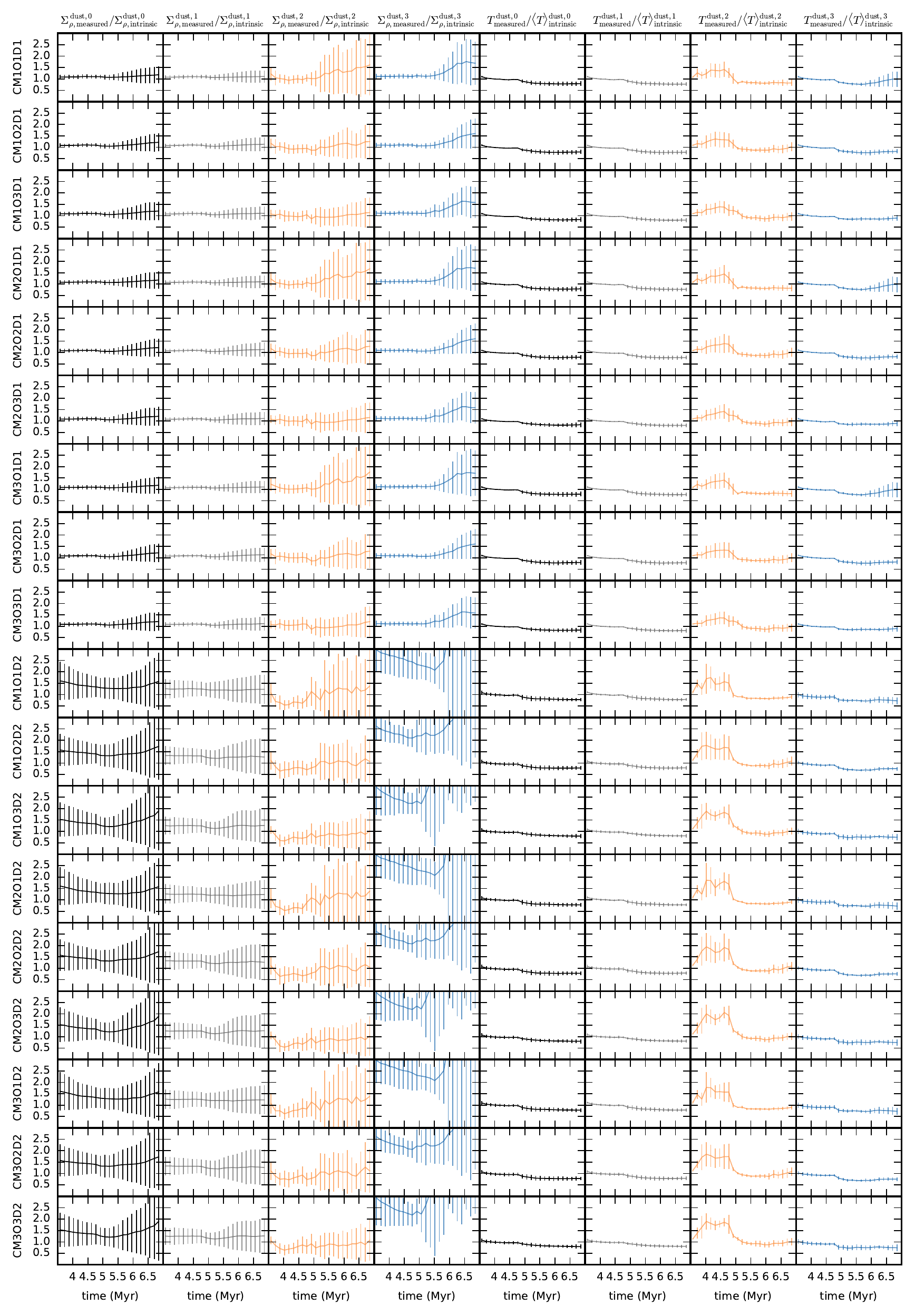}
  \caption{\label{C5:Fig:errors} Evolution of the fraction of measured to intrinsic value for the different orientations (\acs{O1}: xy plane, \acs{O2}: xz plane, \acs{O3}: yz plane), circumstellar set-ups (\acs{CM1}: no refinement, \acs{CM2}: refinement with rotationally flattened envelope, \acs{CM3}: refinement with power-law envelope) and distances (\acs{D1}: \SI{3}{\kpc}, \acs{D2}: \SI{10}{\kpc}) in the background setup \acs{B1}. Displayed is the result for version \acs{R3}, version \acs{R2} will be available as online figure.}
\end{figure*}
%
\subsection{Reliability of Measured Dust Surface Density and Dust Temperature Maps}
\label{C5:Sec:Discussion_Rho_temp}
%
From our detailed study in Section~\ref{C5:Sec:Mgas_Results_SigmaTemp} of the errors of the measured dust surface density $\Sigma_{\textup{measured}}^{\textup{dust}}$ and the measured dust temperature $T_{\textup{measured}}^{\textup{dust}}$ in every pixel, we now give estimates of the reliability.

The median values of the fractions measured to intrinsic values and their $\sigma_{\textup{MAD}}$ are given in Table~\ref{C5:Appendix:median}. The absolute error in percent can be calculated as follows:
\begin{eqnarray}
    \label{C5:Eq:limit}
    \mbox{limits} = - \left[1 - \left(\mbox{MED} \mp \sigma_{\textup{MAD}}\right)\right] \cdot \SI{100}{\percent}.
\end{eqnarray}
We calculate the limits for every value listed in Table~\ref{C5:Appendix:median} for the total cloud as well as for the values of the sub-regions. We plot the results in Figure~\ref{C5:Fig:errors}. We can see that the error for the measured dust temperature is mostly smaller than for the dust surface density. As noted in Section~\ref{C5:Sec:Mgas_Results}, the errors are larger for larger distances. The dust surface density errors grow strongly for the star-formation sites (orange) at later time-steps since more accreting objects are present and more distributed at later time-steps. The errors are even larger at later time-steps for the low-density regions (blue) as the ionized bubble grows over time. This is also the reason why the error is larger at the orientation \acs{O1} where the ionized bubble is most prominent. At early time-steps the dust temperature measurement has larger errors for the star-formation sites. This is due to the large accreting stars present at early time-steps of the \acs{D14} simulations (for more detail see Section~\ref{C5:Sec:Mgas_Results_Mgas}).\\

For a conservative measures of the measured properties, we select for the total cloud and for the sub-regions the largest error limits (Eq.~\ref{C5:Eq:limit}) and suggest to use them as an error estimate of the technique:

\begin{itemize}
    \item total cloud:
    \begin{eqnarray}
        \label{C5:Eq:rho_D1}
        \Sigma_{\rho \textup{, measured}}^{\textup{dust}} & \overset{\textup{D1}}{=} & \Sigma_{\rho \textup{, intrinsic}}^{\textup{dust}}\overset{\textup{R2}}{\big|^{-\SI{22}{\percent}}_{+\SI{71}{\percent}}}\overset{\textup{R3}}{\big|^{-\SI{21}{\percent}}_{+\SI{62}{\percent}}}\\
\label{C5:Eq:rho_D2}
        \Sigma_{\rho \textup{, measured}}^{\textup{dust}} & \overset{\textup{D2}}{=} & \Sigma_{\rho \textup{, intrinsic}}^{\textup{dust}}\overset{\textup{R2}}{\big|^{-\SI{103}{\percent}}_{+\SI{328}{\percent}}}\overset{\textup{R3}}{\big|^{-\SI{82}{\percent}}_{+\SI{264}{\percent}}}\\
        T_{\textup{measured}}^{\textup{dust}} &\overset{\textup{D1}}{=}& \langle T\rangle_{\textup{intrinsic}}^{\textup{dust}} \overset{\textup{R2}}{\big|^{-\SI{35}{\percent}}_{+\SI{13}{\percent}}}\overset{\textup{R3}}{\big|^{-\SI{33}{\percent}}_{+\SI{13}{\percent}}}\\
        T_{\textup{measured}}^{\textup{dust}} &\overset{\textup{D2}}{=}& \langle T\rangle_{\textup{intrinsic}}^{\textup{dust}} \overset{\textup{R2}}{\big|^{-\SI{43}{\percent}}_{+\SI{17}{\percent}}}\overset{\textup{R3}}{\big|^{-\SI{32}{\percent}}_{+\SI{17}{\percent}}}
    \end{eqnarray}
\end{itemize}
    
    \begin{itemize}
    \item sub-region I: bulk molecular cloud
    \begin{eqnarray}
        \Sigma_{\rho \textup{, measured}}^{\textup{dust}} & \overset{\textup{D1}}{=} & \Sigma_{\rho \textup{, intrinsic}}^{\textup{dust}}\overset{\textup{R2}}{\big|^{-\SI{20}{\percent}}_{+\SI{50}{\percent}}}\overset{\textup{R3}}{\big|^{-\SI{19}{\percent}}_{+\SI{44}{\percent}}}\\
        \Sigma_{\rho \textup{, measured}}^{\textup{dust}} & \overset{\textup{D2}}{=} & \Sigma_{\rho \textup{, intrinsic}}^{\textup{dust}}\overset{\textup{R2}}{\big|^{-\SI{63}{\percent}}_{+\SI{140}{\percent}}}\overset{\textup{R3}}{\big|^{-\SI{53}{\percent}}_{+\SI{106}{\percent}}}\\
        T_{\textup{measured}}^{\textup{dust}} &\overset{\textup{D1}}{=}& \langle T\rangle_{\textup{intrinsic}}^{\textup{dust}} \overset{\textup{R2}}{\big|^{-\SI{35}{\percent}}_{+\SI{13}{\percent}}}\overset{\textup{R3}}{\big|^{-\SI{32}{\percent}}_{+\SI{13}{\percent}}}\\
        T_{\textup{measured}}^{\textup{dust}} &\overset{\textup{D2}}{=}& \langle T\rangle_{\textup{intrinsic}}^{\textup{dust}} \overset{\textup{R2}}{\big|^{-\SI{38}{\percent}}_{+\SI{17}{\percent}}}\overset{\textup{R3}}{\big|^{-\SI{31}{\percent}}_{+\SI{16}{\percent}}}
    \end{eqnarray}
\end{itemize}
    \newpage
\begin{itemize}
    \item sub-region II: star-formation sites
    \begin{eqnarray}
        \Sigma_{\rho \textup{, measured}}^{\textup{dust}} & \overset{\textup{D1}}{=} & \Sigma_{\rho \textup{, intrinsic}}^{\textup{dust}}\overset{\textup{R2}}{\big|^{-\SI{77}{\percent}}_{+\SI{249}{\percent}}}\overset{\textup{R3}}{\big|^{-\SI{80}{\percent}}_{+\SI{226}{\percent}}}\\
        \Sigma_{\rho \textup{, measured}}^{\textup{dust}} & \overset{\textup{D2}}{=} & \Sigma_{\rho \textup{, intrinsic}}^{\textup{dust}}\overset{\textup{R2}}{\big|^{-\SI{134}{\percent}}_{+\SI{356}{\percent}}}\overset{\textup{R3}}{\big|^{-\SI{127}{\percent}}_{+\SI{205}{\percent}}}\\
        T_{\textup{measured}}^{\textup{dust}} &\overset{\textup{D1}}{=}& \langle T\rangle_{\textup{intrinsic}}^{\textup{dust}} \overset{\textup{R2}}{\big|^{-\SI{38}{\percent}}_{+\SI{56}{\percent}}}\overset{\textup{R3}}{\big|^{-\SI{32}{\percent}}_{+\SI{83}{\percent}}}\\
        T_{\textup{measured}}^{\textup{dust}} &\overset{\textup{D2}}{=}& \langle T\rangle_{\textup{intrinsic}}^{\textup{dust}} \overset{\textup{R2}}{\big|^{-\SI{47}{\percent}}_{+\SI{62}{\percent}}}\overset{\textup{R3}}{\big|^{-\SI{26}{\percent}}_{+\SI{161}{\percent}}}
    \end{eqnarray}
\end{itemize}
    
\begin{itemize}
    \item sub-region III: low-density regions
    \begin{eqnarray}
        \Sigma_{\rho \textup{, measured}}^{\textup{dust}} & \overset{\textup{D1}}{=} & \Sigma_{\rho \textup{, intrinsic}}^{\textup{dust}}\overset{\textup{R2}}{\big|^{-\SI{33}{\percent}}_{+\SI{220}{\percent}}}\overset{\textup{R3}}{\big|^{-\SI{31}{\percent}}_{+\SI{177}{\percent}}}\\
        \Sigma_{\rho \textup{, measured}}^{\textup{dust}} & \overset{\textup{D2}}{=} & \Sigma_{\rho \textup{, intrinsic}}^{\textup{dust}}\overset{\textup{R2}}{\big|^{-\SI{515}{\percent}}_{+\SI{5422}{\percent}}}\overset{\textup{R3}}{\big|^{-\SI{430}{\percent}}_{+\SI{3990}{\percent}}}\\
        T_{\textup{measured}}^{\textup{dust}} &\overset{\textup{D1}}{=}& \langle T\rangle_{\textup{intrinsic}}^{\textup{dust}} \overset{\textup{R2}}{\big|^{-\SI{42}{\percent}}_{+\SI{23}{\percent}}}\overset{\textup{R3}}{\big|^{-\SI{36}{\percent}}_{+\SI{32}{\percent}}}\\
        T_{\textup{measured}}^{\textup{dust}} &\overset{\textup{D2}}{=}& \langle T\rangle_{\textup{intrinsic}}^{\textup{dust}} \overset{\textup{R2}}{\big|^{-\SI{51}{\percent}}_{+\SI{7}{\percent}}}\overset{\textup{R3}}{\big|^{-\SI{43}{\percent}}_{+\SI{12}{\percent}}}
    \end{eqnarray}
\end{itemize}
The \acs{R2} values listed above are $\chi^2$-corrected.\\

The findings above suggest that method \acs{R3} produce smaller errors for an individual pixel.

\subsection{Reliability of Measured Gas Mass across the Scales}
\label{C5:Sec:Discussion_distances}
In Section~\ref{C5:Sec:Mgas_Results_Mgas}, we showed that the gas mass estimates from modified blackbody fitting work well when there is no background at all or no high background present (which needs to be removed) for objects within a \SIrange{3}{10}{\kpc} distance.

This is also true for closer regions, especially for nearby clouds within \SI{1}{\kpc}. They are statistically not necessarily in positions which fall in the projected Galactic midplane (e.\,g.~above or below the midplane, or in the outer disk opposite the Galactic center). Further, the spatial resolution is better and fewer pixels are contaminated by the "break down" or "under/over-estimation" of the algorithm as stars which are causing the non-constant temperature along the line of sight contaminate relatively fewer pixels of the image.

As pointed out in Section~\ref{C5:Sec:results_B2}, for intermediate-mass regions in the projected Galactic plane, disentangling a specific region from the background is a limiting factor for the success of the technique, but only if we are solely interested in the mass of the star-forming region excluding the mass of the background from the Galactic plane. Otherwise the technique should work better, because then background does not needs to be removed. From this study, we know that intermediate mass regions ($<\SI{e4}{\Msun}$) are lost to the SPIRE background emission. It needs to be tested in future work if simulated star-forming regions above $\SI{e5}{\Msun}$ produce enough significant features in the SPIRE bands to create distinctive features in the far-infrared emission of the Milky Way's midplane.

It remains unclear whether for distant galaxies, where a pixel contains an even larger volume of material, the assumption of a constant temperature along the line of sight still holds for the bulk of mass. This raises interesting questions about the reliability of the technique for extragalactic regimes where unresolved regions are confined to one pixel. As pointed out in Section~\ref{C5:Sec:Mgas_Results_Mgas} $\chi^2$-analysis cannot be used improve the measurement since it is unclear wether the pixels are overestimated or underestimated and additionally because of the large distances too much data/mass is lost due to the large physical size of the pixels.

In Section~\ref{C5:Sec:Mgas_Results_Mgas} we find that the maximum error for distances up to \SI{10}{\kpc} lies between \SI{-7}{\percent} and \SI[retain-explicit-plus]{+20}{\percent} for the \acs{R2} total gas mass measurement and between \SI{-13}{\percent} and \SI[retain-explicit-plus]{+9}{\percent} for the \acs{R3} total gas mass measurement. Keeping in mind the error spread of the individual pixels of the surface density map (see Section~\ref{C5:Sec:Mgas_Results_SigmaTemp} and Section~\ref{C5:Sec:Discussion_Rho_temp}), a very conservative maximum error span can be derived by adapting the same error margins for the total gas mass measurements from the pixel-by-pixel analysis shown in Eq.~\ref{C5:Eq:rho_D1} and Eq.~\ref{C5:Eq:rho_D2}. However, these very conservative error margins will not represent the expected error spread as displayed in Figure~\ref{C5:Fig:Mgas}, where errors are dependent of distance, feedback and resolution version. Below we give the absolute errors from the gas mass measurements (showed in Figure~\ref{C5:Fig:Mgas}) for all time-steps (all), pre-feedback time-steps (co) and high-mass stellar feedback time-steps (fb) for the two resolution version \acs{R2} and \acs{R3} and the two distances \acs{D1} and \acs{D2}. The \acs{R2} values listed below are $\chi^2$-corrected. The first value from the top to bottom is the minimum absolute error, the median error and the maximum absolute error:

    \begin{eqnarray}
        \label{C5:Eq:Mgas_R2_D1}
        M_{\textup{measured}}^{\textup{gas}} &\overset{\textup{D1}}{\underset{\textup{R2}}{=}}&  M_{\textup{intrinsic}}^{\textup{gas}}\  \begin{array}{|c}\scriptsize\textup{all}\\{\scriptsize-\SI{7}{\percent}}\\{\scriptsize+\SI{0}{\percent}}\\{\scriptsize+\SI{14}{\percent}}\end{array}\  \begin{array}{|c}\scriptsize\textup{co}\\{\scriptsize+\SI{4}{\percent}}\\{\scriptsize+\SI{7}{\percent}}\\{\scriptsize+\SI{14}{\percent}}\end{array}\ 
\begin{array}{|c}\scriptsize\textup{fb}\\{\scriptsize-\SI{7}{\percent}}\\{\scriptsize-\SI{3}{\percent}}\\{\scriptsize+\SI{4}{\percent}}\end{array}\\
        \label{C5:Eq:Mgas_R2_D2}
        M_{\textup{measured}}^{\textup{gas}} &\overset{\textup{D2}}{\underset{\textup{R2}}{=}}&  M_{\textup{intrinsic}}^{\textup{gas}}\  \begin{array}{|c}\scriptsize\textup{all}\\{\scriptsize-\SI{5}{\percent}}\\{\scriptsize+\SI{3}{\percent}}\\{\scriptsize+\SI{20}{\percent}}\end{array}\  \begin{array}{|c}\scriptsize\textup{co}\\{\scriptsize+\SI{6}{\percent}}\\{\scriptsize+\SI{11}{\percent}}\\{\scriptsize+\SI{20}{\percent}}\end{array}\ 
\begin{array}{|c}\scriptsize\textup{fb}\\{\scriptsize-\SI{5}{\percent}}\\{\scriptsize+\SI{0}{\percent}}\\{\scriptsize+\SI{6}{\percent}}\end{array}\\
        \label{C5:Eq:Mgas_R3_D1}
        M_{\textup{measured}}^{\textup{gas}} &\overset{\textup{D1}}{\underset{\textup{R3}}{=}}&  M_{\textup{intrinsic}}^{\textup{gas}}\  \begin{array}{|c}\scriptsize\textup{all}\\{\scriptsize-\SI{10}{\percent}}\\{\scriptsize-\SI{5}{\percent}}\\{\scriptsize+\SI{9}{\percent}}\end{array}\  \begin{array}{|c}\scriptsize\textup{co}\\{\scriptsize+\SI{3}{\percent}}\\{\scriptsize+\SI{5}{\percent}}\\{\scriptsize+\SI{9}{\percent}}\end{array}\ \ 
\begin{array}{|c}\scriptsize\textup{fb}\\{\scriptsize-\SI{10}{\percent}}\\{\scriptsize-\SI{7}{\percent}}\\{\scriptsize+\SI{0}{\percent}}\end{array}\\
        \label{C5:Eq:Mgas_R3_D2}
        M_{\textup{measured}}^{\textup{gas}} &\overset{\textup{D2}}{\underset{\textup{R3}}{=}}&  M_{\textup{intrinsic}}^{\textup{gas}}\  \begin{array}{|c}\scriptsize\textup{all}\\{\scriptsize-\SI{13}{\percent}}\\{\scriptsize-\SI{9}{\percent}}\\{\scriptsize+\SI{6}{\percent}}\end{array}\  \begin{array}{|c}\scriptsize\textup{co}\\{\scriptsize-\SI{1}{\percent}}\\{\scriptsize+\SI{2}{\percent}}\\{\scriptsize+\SI{6}{\percent}}\end{array}\ \ 
\begin{array}{|c}\scriptsize\textup{fb}\\{\scriptsize-\SI{13}{\percent}}\\{\scriptsize-\SI{10}{\percent}}\\{\scriptsize-\SI{2}{\percent}}\end{array}
    \end{eqnarray}

We can see that the median error at all time-steps is lower for the \acs{R2} than for the \acs{R3} method. Same goes for the time-steps with high-mass stellar feedback, while the median error for the pre-feedback phase is higher for \acs{R2}. However, as pointed out in Section~\ref{C5:Sec:Mgas_Results_Mgas} the pre-feedback time-steps remain questionable due to the unorthodox setup of the \acs{D14} simulations. Therefore, the \acs{R2} technique appears to produce more robust results for the total gas mass measurements.

Additionally, we have to keep in mind for the version \acs{R2} as for \acs{R3} that once a realistic background is included, the errors will grow, but these errors will then be due to overestimation or underestimation of the non-uniform background (see Section~\ref{C5:Sec:results_BB}).

Whether the technique still works for more complex systems such as distant galaxies, where a pixel combines several \si{pc} or even \si{kpc}, and whether the assumption of a constant temperature along the line of sight still holds for the bulk of the mass. This raises interesting questions about the reliability of the technique for extragalactic regimes where unresolved regions are confined to one pixel. There even the $\chi^2$-analysis cannot be used too excessively because then too much data/mass is lost due to large physical pixel sizes.

\section{Summary}
\label{C5:Sec:Summary}
In this paper \citepalias{KDR2a:inprep}, we explored a commonly used observational measurement technique to infer the gas and dust properties on $\sim$5800 synthetic observations of a simulated \SI{30}{\pc} wide star-forming region at different evolutionary time-steps, orientations, distances and different circumstellar radiative transfer set-ups. We tested the reliability of modified blackbody fitting to estimate dust temperature and dust surface density maps and to recover the total gas mass. We found in \citetalias{KDR1:inprep} \citep{KDR1:inprep} that in regions of high background of the \emph{Herschel} bands, it is not possible to disentangle the studied intermediate-mass star-forming regions ($\sim$ \SI{e4}{\Msun}) from the background. With low background (e.\,g.~off the Galactic plane), this technique should work much better. The technique works well for synthetic observations without a real background from the Galactic plane when measuring the total gas mass of the cloud when assuming the correct distance and interstellar extinction.

When using images which have been convolved to the largest common beam size (version \acs{R3}), we measure the total gas mass with median errors up to \SI[retain-explicit-plus]{-9}{\percent} from the actual value. On the contrary, when using images which have been only scaled to the larges common pixel size (version \acs{R2}), we measure the total gas mass median errors up to \SI[retain-explicit-plus]{+3}{\percent} from the actual value. While for the latter technique the errors is lower for time-steps with high-mass stellar feedback present, for the \acs{R3} technique the error grows in the high-mass stellar feedback phase (underestimation). For both techniques the errors increase with distance.

Below we present the absolute error span of the total gas mass measurements for all time-steps. The first value from the top to bottom is the minimum absolute error, the median error and the maximum absolute error:
    \begin{eqnarray}
        M_{\textup{measured}}^{\textup{gas}} &\overset{\SI{3}{\kpc}}{\underset{\textup{R2}}{=}}&  M_{\textup{intrinsic}}^{\textup{gas}}\  \begin{array}{|c}{\scriptsize-\SI{7}{\percent}}\\{\scriptsize+\SI{0}{\percent}}\\{\scriptsize+\SI{14}{\percent}}\end{array}\\
        M_{\textup{measured}}^{\textup{gas}} &\overset{\SI{10}{\kpc}}{\underset{\textup{R2}}{=}}&  M_{\textup{intrinsic}}^{\textup{gas}}\  \begin{array}{|c}{\scriptsize-\SI{5}{\percent}}\\{\scriptsize+\SI{3}{\percent}}\\{\scriptsize+\SI{20}{\percent}}\end{array}\\
        M_{\textup{measured}}^{\textup{gas}} &\overset{\SI{3}{\kpc}}{\underset{\textup{R3}}{=}}&  M_{\textup{intrinsic}}^{\textup{gas}}\  \begin{array}{|c}{\scriptsize-\SI{10}{\percent}}\\{\scriptsize-\SI{5}{\percent}}\\{\scriptsize+\SI{9}{\percent}}\end{array}\\
        M_{\textup{measured}}^{\textup{gas}} &\overset{\SI{10}{\kpc}}{\underset{\textup{R3}}{=}}&  M_{\textup{intrinsic}}^{\textup{gas}}\  \begin{array}{|c}{\scriptsize-\SI{13}{\percent}}\\{\scriptsize-\SI{9}{\percent}}\\{\scriptsize+\SI{6}{\percent}}\end{array}
    \end{eqnarray}
We find that the total gas mass measurements have lower median errors for method \acs{R2} than for \acs{R3} especially for larger distances.

We emphasize that when the initial images are convolved to the largest common beam size information is lost and contaminated pixels are washed out over larger areas, making the technique less reliable conceptually although the absolute errors might appear a little smaller. Techniques which keep the initial beam resolution  (e.\,g.~\citealt{Marsh:2015}) minimize the loss of information and likely produce more reliable results. Further, the \acs{R3} measurement are most likely underestimated in the presence of high-mass stellar feedback and $\chi^2-$values can not be trusted by conception. Therefore, masking the pixels with unreliable fits ($\chi^2$-analysis) should only be applied when the images have not been convolved to the largest beam size.

When analyzing the dust surface density and dust temperature measurements for every pixel we find that the error from measured to intrinsic value rises for cells where there is a large gradient in temperature and density along the line of sight and spatially. This is especially the case for regions closer than \SI{1}{\pc} to the star-formation sites and even more for low-density regions below \SI{e-6}{\gcm} with derived values as large as \num{50}-times to \num{70}-times above the actual value. Generally we find that the \acs{R3} measurements produce slightly more accurate individual pixel values.

We provide our measured gas masses and the actual value as well as the median values and error spread of the dust surface density and dust temperature pixel analysis in Table~\ref{C5:Appendix:median} in Appendix~\ref{C5:Appendix}. Moreover, the dust temperature and dust surface density maps are provided in the online material of Appendix~\ref{C5:Appendix}. In follow-up papers (\citetalias{KDR2b:inprep}: \citealt{KDR2b:inprep}, \citetalias{KDR3:inprep}: \citeauthor[in prep.]{KDR3:inprep}), we will examine the reliability of different diffuse star-formation dust tracers and direct counting techniques to calculate the \ac{SFR}.

\section{Acknowledgements}
We thank the referee for a constructive report that helped us improve the clarity and the strength of the results presented in our paper.
This work was carried out in the Max Planck Research Group \textit{Star formation throughout the Milky Way Galaxy} at the Max Planck Institute for Astronomy. C.K. is a fellow of the International Max Planck Research School for Astronomy and Cosmic Physics (IMPRS) at the University of Heidelberg, Germany and acknowledges support. C. K. acknowledges support from STFC grant ST/M001296/1.
J. E. D. was supported by the DFG cluster of excellence \textit{Origin and Structure of the Universe}. 
This research made use of Astropy, a community-developed core Python package for Astronomy \citep{Astropy:2013}, matplotlib, a Python plotting library \citep{Hunter:2007}, Scipy, an open source scientific computing tool \citep{Scipy}, the NumPy package \citep{NumPy} and IPython, an interactive Python application \citep{IPython}.

\bibliographystyle{latex_apj}
\bibliography{latex_ref}

\appendix

\section{Measurements and Online Material}
\label{C5:Appendix}

In the online material, we provide 414 dust surface density, dust temperature and $\chi^2$ maps (\acs{FITS} files) of the 3 circumstellar material set-ups, 23 time-steps, 3 viewing angles, 2 distances of the synthetic star-forming region. The \acs{FITS} files will also be presented in the online material. The \acs{R2} and \acs{R3} dust surface density and dust temperature maps can be accessed from:

\begin{center}
 R2: \url{http://dx.doi.org/10.5281/zenodo.31294}\\
 \vspace{0.2cm}
 R3: \url{http://dx.doi.org/10.5281/zenodo.56424}
 \end{center}
The filename of the maps have the extension \verb|_B1_MBBF.fits| and \verb|_B1_MBBF_R3.fits| for the respective versions (\acs{R2} and \acs{R3}) and are constructed of the combinations of abbreviations which resemble methods from \citetalias{KDR1:inprep} and are separated by underscores:
\begin{lstlisting}[numbers=none,basicstyle=\normalsize]
both_I122_s3_p3_voronoi_DraineLiPAH_c1_DT2_CM1_O1_D1_B1_MBBF.fits
\end{lstlisting}
For more details about the different abbreviations in the filenames, see \citetalias{KDR1:inprep}. The individual \acs{FITS} files contain three maps: dust surface density (\si{\gram\per\centi\meter\squared}), dust temperature (\si{\kelvin}) and $\chi^2$ maps. \\

In the online figures we present the \acs{R2} counterparts to \acs{R3} versions of Figure~\ref{C5:Fig:surface_density}, Figure~\ref{C5:Fig:temperature}, Figure~\ref{C5:Fig:histogram}, Figure~\ref{C5:Fig:maps}, Figure~\ref{C5:Fig:scatter_D1}, Figure~\ref{C5:Fig:scatter_D2} and Figure~\ref{C5:Fig:errors}.\\

In this appendix, we list the real and the measured gas mass and medians/errors of the dust surface density and dust temperature measurements for the different combinations of circumstellar set-ups (\acs{CM1}: no refinement, \acs{CM2}: refinement with rotationally flattened envelope, \acs{CM3}: refinement with power-law envelope), orientations (\acs{O1}: xy plane, \acs{O2}: xz plane, \acs{O3}: yz plane), distances (\acs{D1}: \SI{3}{\kpc}, \acs{D2}: \SI{10}{\kpc}), resolution version (\acs{R2}: scaled to pixel size of \ac{SPIRE} \SI{500}{\microns}, \acs{R3}: convolved to beam size and scaled to pixel size of \ac{SPIRE} \SI{500}{\microns}) and sub-regions (0: total cloud, 1: bulk molecular cloud, 2: star-formation sites, 3: low-density regions).

  \begingroup
  \tiny \renewcommand{\arraystretch}{0.8} \setlength{\tabcolsep}{2.3pt}
\begin{longtable}{lccc cccc ccccc}
\caption{\label{C5:Appendix:median}\normalsize Intrinsic and Measured Properties for Background \acs{B1}}\\
\hline\\[-13pt]
\hline\\[-5pt]   
&&&& &&&& &&&&\\
run & \multicolumn{1}{c}{time} & \multicolumn{1}{c}{sorter} &
\multicolumn{1}{c}{$M_{\textup{gas}}^{\textup{sim}}$} & \multicolumn{1}{c}{$M_{\textup{gas}}^{\textup{B1}}$} &
\multicolumn{4}{c}{$\Sigma_{\rho \textup{, measured}}^{\textup{\,dust}} / \Sigma_{\rho \textup{, intrinsic}}^{\textup{dust}}$} & \multicolumn{4}{c}{$T_{\textup{ measured}}^{\textup{\,dust}} / \langle T\rangle_{\textup{intrinsic}}^{\textup{dust}}$}\\

(ID) & \multicolumn{1}{c}{(\si{\Myr})} &  &  \multicolumn{1}{c}{(\si{\Msun})} &  \multicolumn{1}{c}{(\si{\Msun})} 
&\multicolumn{4}{c}{(median $\pm\ \sigma_{\textup{MAD}}$)} 
&\multicolumn{4}{c}{(median $\pm\ \sigma_{\textup{MAD}}$)}\\[6pt]
 				\hline\\
\multicolumn{3}{l}{sub-region in cloud}&0&0&0&1&2&3&0&1&2&3\\[6pt]
 				\hline\\[-5pt]
\endfirsthead
\endhead    
 &&&& &&&& &&&&\\[-5pt]
024 & 3.576 & CM1O1D1R2 &9873 & 10313 &\num{1.09\pm0.11} & \num{1.08\pm0.10} & \num{1.32\pm0.38} & \num{1.13\pm0.14} & \num{1.09\pm0.04} & \num{1.09\pm0.03} & \num{1.05\pm0.07} & \num{1.09\pm0.06}\\
025 & 3.725 & CM1O1D1R2 &9825 & 10330 &\num{1.09\pm0.11} & \num{1.08\pm0.10} & \num{1.29\pm0.39} & \num{1.12\pm0.14} & \num{1.03\pm0.03} & \num{1.03\pm0.03} & \num{1.19\pm0.15} & \num{1.02\pm0.05}\\
\ ...&...&...&...&...&...&...&...&...&...&...&...&...\\
&&&& &&&& &&&&\\
\hline\\[-5pt]
\end{longtable}
\endgroup
\vspace*{-0.3cm}
\footnotesize{The full table will be provided in the online material including the $\chi^2$-corrected values.}

\begin{acronym}[ATLASGAL]
\acro{2MASS}{Two Micron All-Sky Survey}
\acro{AGB}{Asymptotic Giant Branch}
\acro{ALMA}{Atacama Large Millimeter/Submillimeter Array}
\acro{AMR}{Adaptive Mesh Refinement}
\acro{ATLASGAL}{APEX Telescope Large Area Survey of the Galaxy}
\acro{BGPS}{Bolocam Galactic Plane Survey}
\acro{B1}{realistic synthetic observation not combined with real background}
\acro{B2}{realistic synthetic observation combined with real background}
\acro{c1}{sample clipping of only neutral particles within a box of \SI{30}{\pc}}
\acro{c2}{sample clipping of \acs{c1} particles and for a certain threshold temperature}
\acro{c2d}{Cores to Disks Legacy}
\acro{CASA}{Common Astronomy Software Applications package}
\acro{cm}{centimeter}
\acro{CMF}{core mass function}
\acro{CMZ}{central molecular zone}
\acro{D1}{distance at \SI{3}{\kpc}}
\acro{D14}{\acs{SPH} simulations performed by Jim Dale and collaborators \citep{DaleI:2011,DaleIoni:2012,DaleIoni:2013,DaleWind:2013,DaleBoth:2014}}
\acro{D2}{distance at \SI{10}{\kpc}}
\acro{DT1}{temperature coupling of radiative transfer \& hydrodynamical temperature}
\acro{DT2}{temperature coupling of the radiative transfer \& isothermal temperature}
\acro{DT3}{no temperature coupling of the radiative transfer temperature}
\acro{e1}{using the \cite{Ulrich:1976} envelope profile to extrapolate the envelope inwards}
\acro{e2}{using the \cite{Ulrich:1976} envelope profile with suppressed singularity to extrapolate the envelope inwards}
\acro{e3}{using a power-law envelope profile to extrapolate the envelope inwards}
\acro{EOS}{equation of state}
\acro{FIR}{far-infrared}
\acro{FITS}{Flexible Image Transport System}
\acro{FWHM}{full-width at half-maximum}
\acro{GLIMPSE}{Galactic Legacy Infrared Mid-Plane Survey Extraordinaire}
\acro{GMC}{Giant Molecular Clouds}
\acro{Hi-GAL}{\emph{Herschel} Infrared Galactic Plane Survey}
\acro{HST}{\emph{Hubble} Space Telescope}
\acro{HWHM}{half-width at half-maximum}
\acro{IMF}{initial mass function}
\acro{IR}{infrared}
\acro{IRAC}{Infrared Array Camera}
\acro{IRAS}{Infrared Astronomical Satellite}
\acro{ISM}{interstellar medium}
\acro{$K$}{K band}
\acro{LTE}{local thermodynamical equilibrium}
\acro{MAD}{median absolute deviation}
\acro{MIPS}{Multiband Imaging Photometer for \emph{Spitzer}}
\acro{MIPSGAL}{\acs{MIPS} Galactic Plane Survey}
\acro{MIR}{mid-infrared}
\acro{MS}{main-sequence}
\acro{mm}{millimeter}
\acro{NASA}{National Aeronautics and Space Administration}
\acro{NIR}{near-infrared}
\acro{N-PDF}{column density \acs{PDF}}
 \acro{O1}{xy plane}
 \acro{O2}{xz plane}
 \acro{O3}{yz plane}
  
\acro{p1}{parameter evaluation version from \acs{SPH} kernel function}
\acro{p2}{parameter evaluation version from \acs{SPH} splitted kernel distribution}
\acro{p3}{parameter evaluation version from \acs{SPH} random distribution}
\acro{PACS}{Photoconductor Array Camera and Spectrometer}
\acro{PAH}{polycyclic aromatic hydrocarbon}
\acro{PDF}{probability distribution function}
\acro{PDR}{Photon Dominated Region}
\acro{PSF}{point-spread-function}
\acro{px}{one of the parameter evaluation version \acs{p1}, \acs{p2}, \acs{p3}}
\acro{R1}{images with band specific pixel resolution and beam size}
\acro{R2}{scaled pixels to largest common pixel size of selected bands}
\acro{R3}{scaled pixels to largest common pixel size and convolved to largest common beam size of selected bands}
\acro{RGB}{red, green and blue}
\acro{CM1}{circumstellar setup with background density and sink mass as stellar mass}
\acro{CM2}{circumstellar setup by a toy model with \acs{e2} envelope superposition density and corrected stellar mass, protoplanetary disk and envelope cavity}
\acro{CM3}{circumstellar setup by a toy model with \acs{e3} envelope superposition density and corrected stellar mass, protoplanetary disk and envelope cavity}
\acro{s1}{Voronoi site placement version at \acs{SPH} particle position}
\acro{s2}{Voronoi site placement version as \acs{s1} including sites at sink particles}
\acro{s3}{Voronoi site placement version as \acs{s2} including circumstellar sites}
\acro{SAO}{Smithsonian Astrophysical Observatory}
\acro{SED}{spectral energy distribution}
\acro{SFE}{star-formation efficiency}
\acro{SFR}{star-formation rate}
\acro{SFR24}{technique to measure the \acs{SFR} using the \SI{24}{\microns} tracer}
\acro{SFR70}{technique to measure the \acs{SFR} using the \SI{70}{\microns} tracer}
\acro{SFRIR}{technique to measure the \acs{SFR} using the total infrared tracer}
\acro{Sgr}{Sagittarius}
\acro{SMA}{Submillimeter Array}
\acro{SPH}{smoothed particle hydrodynamics}
\acro{SPIRE}{Spectral and Photometric Imaging Receiver}
\acro{sub-mm}{sub-millimeter}
\acro{UKIDSS}{UKIRT Infrared Deep-Sky Survey}
\acro{UKIRT}{UK Infrared Telescope}
\acro{UV}{ultra-violet}
\acro{WFCAM}{\acs{UKIRT} Wide Field Camera}
\acro{WISE}{Wide-field Infrared Survey Explorer}
\acro{YSO}{young stellar object}
\end{acronym}

\end{document}